\documentclass[%
 prb,
 amsmath,amssymb,
 reprint,%
]{revtex4-2}
\def\cbcn{\text{C}_\text{B}\text{C}_\text{N}}
\def\nbvn{\text{N}_\text{B}\text{V}_\text{N}}
\def\sns2{\text{SnS}_\text{2}}
\def\mos2{\text{MoS}_\text{2}}
\def\gw{\text{G}_0\text{W}_0}

\usepackage{graphicx}% Include figure files
\usepackage{dcolumn}% Align table columns on decimal point
\usepackage{bm}% bold math
\usepackage{amsmath,amsfonts} %use amsfont for identity matrix
\usepackage{todonotes}
\usepackage{hyperref}
\usepackage{array}
\usepackage{color}
\usepackage{caption}
\usepackage{physics}
\usepackage{soul} % for strikeout text
\DeclareUnicodeCharacter{0301}{\'{e}}

\usepackage{comment}
\definecolor{kc}{rgb}{1,0.5,0}
\definecolor{del}{rgb}{0.66,0.66,0.66} %deleted part
\definecolor{sz}{rgb}{0.03,0.27,0.49} %Shimin's correction

\begin{document}
\preprint{AIP/123-QED}

\title{Effect of Environmental Screening and Strain on Optoelectronic Properties of Two-Dimensional Quantum Defects}% Force line breaks with \\

\author{Shimin Zhang}
\affiliation{Department of Physics, University of California, Santa Cruz, CA, 95064, USA}
\author{Kejun Li}
\affiliation{Department of Physics, University of California, Santa Cruz, CA, 95064, USA}
\author{Chunhao Guo}
\affiliation{Department of Chemistry and Biochemistry, University of California, Santa Cruz, CA, 95064, USA}
\author{Yuan Ping}
\email{yuanping@ucsc.edu}
\affiliation{Department of Chemistry and Biochemistry, University of California, Santa Cruz, CA, 95064, USA}

\date{\today}

\begin{abstract}

Point defects in hexagonal boron nitride (hBN) are promising candidates as single-photon emitters (SPEs) in nanophotonics and quantum information applications. The precise control of SPEs requires in-depth understanding of their optoelectronic properties. However, how the surrounding environment of host materials, including number of layers, substrates, and strain, influences SPEs has not been fully understood. In this work, we study the dielectric screening effect due to the number of layers and substrates, and the strain effect on the optical properties of carbon dimer and nitrogen vacancy defects in hBN from first-principles many-body perturbation theory. We report that the environmental screening causes lowering of the GW gap and exciton binding energy, leading to nearly constant optical excitation energy and exciton radiative lifetime. We explain the results with an analytical model starting from the BSE Hamiltonian with Wannier basis. 
We also show that optical properties of quantum defects are largely tunable by strain with highly anisotropic response, in good agreement with experimental  measurements. Our work clarifies the effect of environmental screening and strain on optoelectronic properties of quantum defects in two-dimensional insulators, facilitating future applications of SPEs and spin qubits in low-dimensional systems.
\end{abstract}

\maketitle
%______________________________________________________________________________________
%______________________INTRODUCTION____________________________________________________
%______________________________________________________________________________________
\section{Introduction}
Point defects in two-dimensional (2D) materials emerge to possess outstanding quantum properties such as stable single-photon emission, and have been exploited as spin quantum bits (qubits) for quantum information technologies~\cite{aharonovich2017quantum,liu20192d}. The single photon emitters (SPEs) in 2D materials are highly stable and tunable~\cite{liu20192d,wolfowicz2021quantum,yim2020polarization}, and in particular, their optical activation can be spatially controlled and tuned by strain~\cite{liGiantShiftStrain2020,mendelsonStrainInducedModificationOptical2020}, emphasizing the great potential of SPEs in 2D materials. 
%%%%%%%%%\textcolor{red}{YP: deterministic defect creation}\textcolor{red}{The high tunability and stability of SPEs on sharp areas like 2D flakes edge, wrinkle and grain boundary provide pathways to control their location and concentration, and make engineering possible~\cite{liu20192d,wolfowicz2021quantum,yim2020polarization}.
%Additionally, spin defects as SPEs in 2D hexagonal boron nitride (hBN)~\cite{cassabois2016hexagonal} can possess deep-level spin states or form defect-bound exciton, making themselves potential ideal photonic qubit generators~\cite{liu20192d}.

The defect candidates with promising quantum properties exhibit deep-level states or form defect-bound exciton. A large number of defects in hBN have been proposed since the report by Tran et al.~\cite{tran2016quantummono}. So far, spin defect $\mathrm{V_B^-}$ has been unambiguously identified from experiment~\cite{gottscholl2020initialization,gottscholl2021room,gao2022nuclear,qian2022unveiling} and theory~\cite{reimers2020photoluminescence,ivady2020ab}. Many of the other defects, whose atomic origins are yet to be determined, were found to be ${\sim}2$ eV and ${\sim}4$ eV SPEs~\cite{grosso2017tunable,mendelson2020identifying,bourrellier2016bright,museur2008defect}. From theoretical predictions,  $\nbvn$~\cite{tran2016quantummono,grosso2017tunable}, boron dangling bonds~\cite{turiansky2019dangling}, $\mathrm{C_BV_N}$~\cite{sajid2020vncb,fischer2021controlled}, and carbon trimers~\cite{jara2021first,li2022carbon,golami2022b,maciaszek2022thermodynamics} were proposed to be defect candidates for the ${\sim}2$ eV SPEs, while $\cbcn$~\cite{mackoit2019carbon}, $\mathrm{C_NO_N}$~\cite{vokhmintsev2019electron}, Stone-Wales defect~\cite{hamdi2020stone}, and carbon ring~\cite{li2022ultraviolet} were for the ${\sim}4$ eV SPEs. 

Among the proposed defect candidates, only partial experimental observations can be explained.
%with the above proposed defect candidates. 
Most importantly, large variations of key physical properties including Zero-Phonon Line (ZPL), photoluminescence (PL) lifetime, and Huang-Rhys (HR) factor~\cite{tran2016robust,li2022carbon} were observed. The physical origin of such variation is undetermined, with only some plausible explanations.  For example, it is speculated that different substrates or sample thickness used in experiments may lead to   variation~\cite{tran2016quantummono,mendelson2020identifying,krecmarova2021extrinsic}. %And the first-principles studies of defects using monolayer, multilayer or bulk hBN models~\cite{ivady2020ab,mackoit2019carbon,jara2021first,li2022carbon,golami2022b,maciaszek2022thermodynamics} has not been fully justified. 
Strain can be another source for the variation, as indicated by the past experimental studies ~\cite{wu2019carrier,grosso2017tunable,dev2020fingerprinting}. Natural strain can be introduced when placing materials on top of substrates. 
%\st{On the other hand, strain has been used as an effective knob for tuning the optical properties of SPEs experimentally, although the underlying mechanism is not completely understood}~\cite{liGiantShiftStrain2020,mendelsonStrainInducedModificationOptical2020}. \textcolor{kc}{K: It seems not necessary to mention the tunability by strain again.}
\\\indent 
Theoretically, first-principles computation has been a powerful tool for identifying and proposing new defects as SPEs and spin qubits in 2D materials~\cite{pingComputationalDesignQuantum2021}. 
However, different structural models including monolayer, 
multilayer or bulk hBN~\cite{ivady2020ab,mackoit2019carbon,jara2021first,li2022carbon,golami2022b,maciaszek2022thermodynamics} were used in different studies, which lead to difficulties of comparing with experiments and comparison among different theoretical studies. Furthermore, the effect of substrates has been mostly examined at the mean-field level by DFT with semilocal functionals, where excitonic effects are not considered~\cite{amblardUniversalPolarisationEnergies2022,wangLayerDependenceDefect2020,wangSubstrateEffectsCharged2019}. 
At last, the effect of strain on optical properties such as absorption spectra, exciton binding energies, and radiative lifetime has not been investigated to our best knowledge. \\\indent
In this work, from first-principles calculations, we investigate the environmental screening effect due to the layer thickness and substrates, as well as the strain effect on the optoelectronic properties of point defects in hBN. In order to pick representative defects for general conclusions, we choose $\cbcn$ as an example of extrinsic substitutional defects and $\nbvn$ as an example of native vacancy defects, both of which were commonly found in hBN and previously proposed to be possible ${\sim}4$ eV and ${\sim}2$ eV SPEs respectively. 
Our results provide an estimation on how sensitive the excitation energy, exciton binding energy, and ZPL are to strain, and we explain their qualitative trends through molecular orbital theory. 
%\st{ZPL strain susceptibility of $\cbcn$ and $\nbvn$ with the consideration of many-body effect and Frank-Condon shift, among which the 2 eV $\nbvn$ shows a reasonable agreement with available experiment.} 
We also provide intuitive and comprehensive understanding of environmental screening effect on defect properties through both first-principles many-body perturbation theory calculations and analytical models. 
%
%analyze the substrate or layer thickness effect on optical properties with analytical models, which provides intuitive understanding on environmental screening effect on defect properties. 
%
%demonstrated robust optical properties of defect SPEs under layer and substrate effect with both first principles calculation and analytical model.

%______________________________________________________________________________________
\section{Computational Method}

The ground state calculations are carried out by density functional theory implemented in QuantumEspresso 
package\cite{giannozziQUANTUMESPRESSOModular2009}, with the Perdew-Burke-Ernzerhof (PBE) exchange correlation functional~\cite{perdewGeneralizedGradientApproximation1996}.  We use the SG15 Optimized Norm-Conserving Vanderbilt (ONCV) 
pseudopotentials~\cite{schlipfOptimizationAlgorithmGeneration2015,hamannOptimizedNormconservingVanderbilt2013} and 
80 Ry wave function kinetic energy cutoff (320 Ry charge density cutoff) for plane wave basis set. 
%the monolayer defect system calculation 
%in substrate and layer effect section %\textcolor{sz}{and 55 Ry for the strain effect calculations in order to reduce the computational cost}. \textcolor{kc}{K: and 55 Ry for the strain effect calculations in order to reduce the computational cost.} \textcolor{del}{Although in 
%strain effect calculation we use 55 Ry energy cutoff in order to speed up the calculation, we make sure the ground state result is consistent with the substrate and layer effect calculation (defect energies difference within  2 meV).}%% 
The defect calculations are performed with $6\times 6$ supercell and $3\times3\times1$ k-point sampling, 
based on our convergence tests in previous studies~\cite{li2022carbon,wu2019carrier,smartIntersystemCrossingExciton2021}.

We perform the many-body perturbation theory calculations with $\text{G}_0\text{W}_0$ 
starting from PBE electronic states for the quasi-particle energies,  and solve the Bethe-Salpeter equation (BSE) for the optical properties~\cite{pingElectronicExcitationsLight2013} with excitonic effects by the Yambo-code~\cite{mariniYamboInitioTool2009b}. 
The GW calculation is carried out with 8 Ry response block size and 1800 energy bands for the 
dielectric matrix and self-energy,  while we use 5 Ry and 80 bands for the BSE kernel and optical spectra calculations.  With the 2D Coulomb truncation technique~\cite{rozziExactCoulombCutoff2006}, the quasi-particle energies are converged within 10 meV at 33.5 a.u. vacuum size. More details on convergence tests can be found in SI Fig.S1 and Fig.S2. 

We then calculate the ZPL by subtracting the Frank-Condon shift $E_{\text{FC}}$ from the BSE excitation energy, where the $ E_{\text{FC}}$ is obtained by the constrained DFT (cDFT) technique~\cite{smartIntersystemCrossingExciton2021}.
By taking the excitation energy and exciton dipole moment from the solution of BSE, we then evaluate radiative lifetime for defects in 2D systems derived from the Fermi's golden rule, with $\tau_R=3\pi\epsilon_0 h^4 c^3/n_D e^2E^3\mu^2$~\cite{wuDimensionalityAnisotropicityDependence2019b,smartIntersystemCrossingExciton2021, smartIntersystemCrossingExciton2021}. %\textcolor{red}{$\tau=3c^3/4E^3\mu^2$. YP: using Eq.9 in Tyler's npj paper}
Here $E$ is the exciton energy, $c$ is the speed of light, $\mu^2$ is the modulus square of exciton dipole moment, and $n_D$ is reflective index, which is one for monolayer hBN.

\begin{comment}
\textcolor{del}{The uniaxial strain here is defined as the stretching ratio of the lattice along certain direction $\epsilon=(l-l_0)/l_0$, where $l_0$ and $l$ are the lattice length before and after applying the strain.
With $\theta$ to be the angle of the in-plane rotation axis with respect to the $C_{2v}$ axis, the strain tensor $\hat{\epsilon}$ components 
can be written as: 
\begin{eqnarray}
   && \epsilon_{xx}=\epsilon \text{cos}^2\theta \nonumber, \epsilon_{yy}=\epsilon sin^2 \theta \nonumber\\
    &&\epsilon_{xy}=\epsilon_{yx}= \epsilon sin\theta cos\theta
\end{eqnarray}}
\end{comment}
%\textcolor{red}{Too much explanation}\textcolor{del}{Due to the weak Van der waal interaction between the 2D materials layers and the highly localized nature defect states, the hybridization of wave function between layers is weak, thus the substrate and layer effect on system electronic and optical properties would be dominated by dielectric screening. To account for the dielectric screening from other layers, we utilize the recently developed sum-up effective polarizability method ($\chi_{\text{eff}}$-sum),which enables the accurate description of substrate effect and largely reduces the computational cost compared to the explicit method.~\cite{guoSubstrateScreeningApproach2020}. The advantage of this method allow us to conduct the pristine layer calculation with single unit cell without the restriction on enforcing lattice match (which will require building a supercell and introducing artificial strain ).}

For the study of layer thickness and substrate effects, we apply our recently-developed sum-up effective polarizability ($\chi_{\text{eff}}$-sum) with the reciprocal-space linear interpolation technique, in order to account for the impact of substrates and multilayers\cite{guoSubstrateScreeningApproach2020,guoSubstrateEffectExcitonic2021}. This method allows us to separate the total interface into two subsystems as substrate pristine layers and defective monolayer, allowing a large saving of computational cost and avoiding artificial strain from enforcing lattice matching at interfaces.

%_______________________________________________________________________________________________
%______________________Result and discussion____________________________________________________
%_______________________________________________________________________________________________
\section{Results and Discussions}
    
%______________Defect properties_______________
%Figure1: CBCN structure
\begin{figure}[h]
    \centering
    \includegraphics[width=\linewidth]{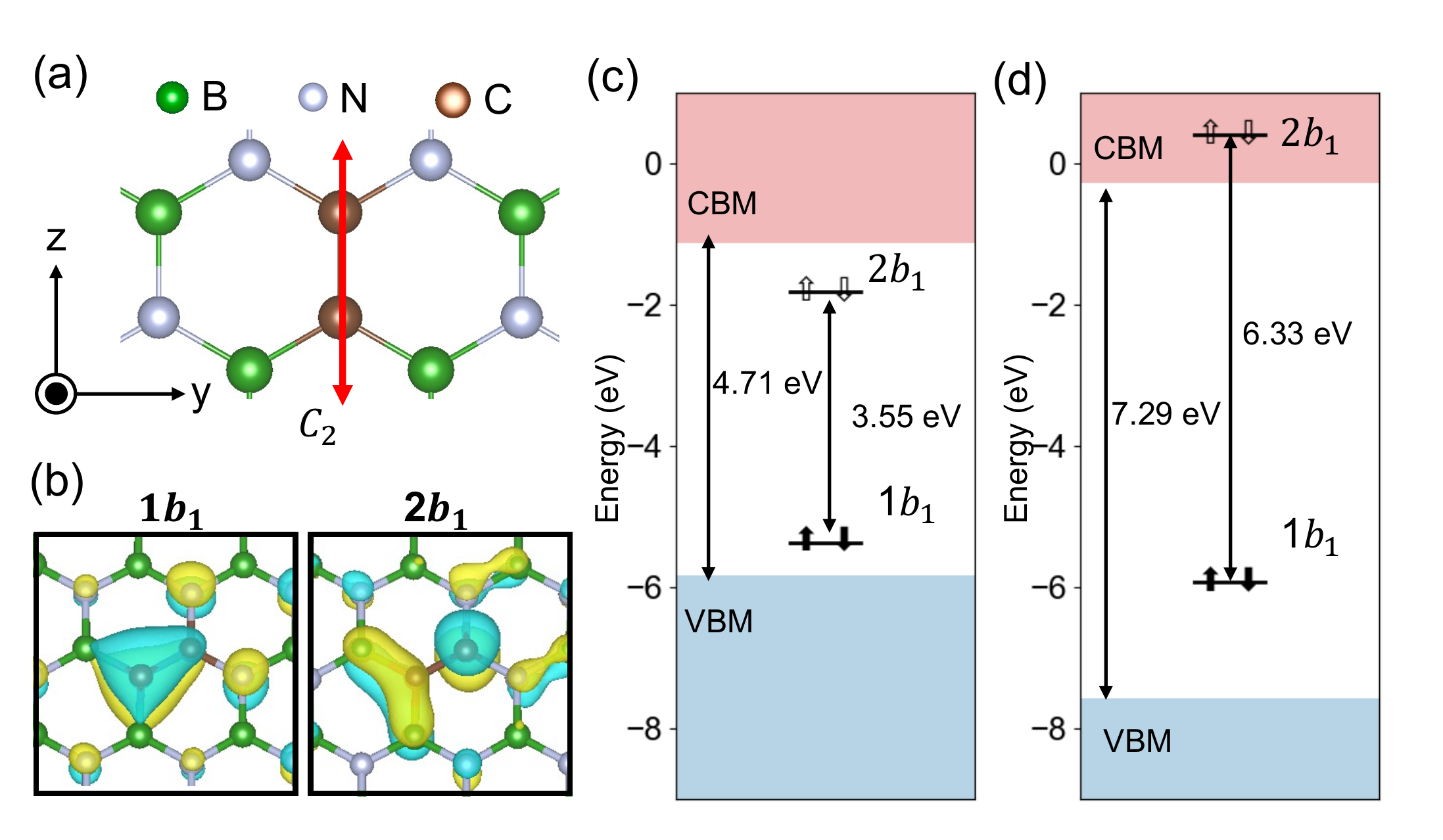}
    \caption{Structural and electronic properties of carbon dimer ($\cbcn$) in hBN. (a) Atomic structure, (b) defect-related wave functions, and the single-particle diagram of ground state at the level of (c) PBE and (d) $\text{G}_0\text{W}_0$@PBE. The zero energy is aligned to the vacuum level, and the defect states are labeled by their wave function symmetry based on the irreducible representation of $C_{2v}$ symmetry group. }
    \label{Fig1}
\end{figure}
%Figure2: NBVN structure
\begin{figure}[h]
    \centering
    \includegraphics[width=\linewidth]{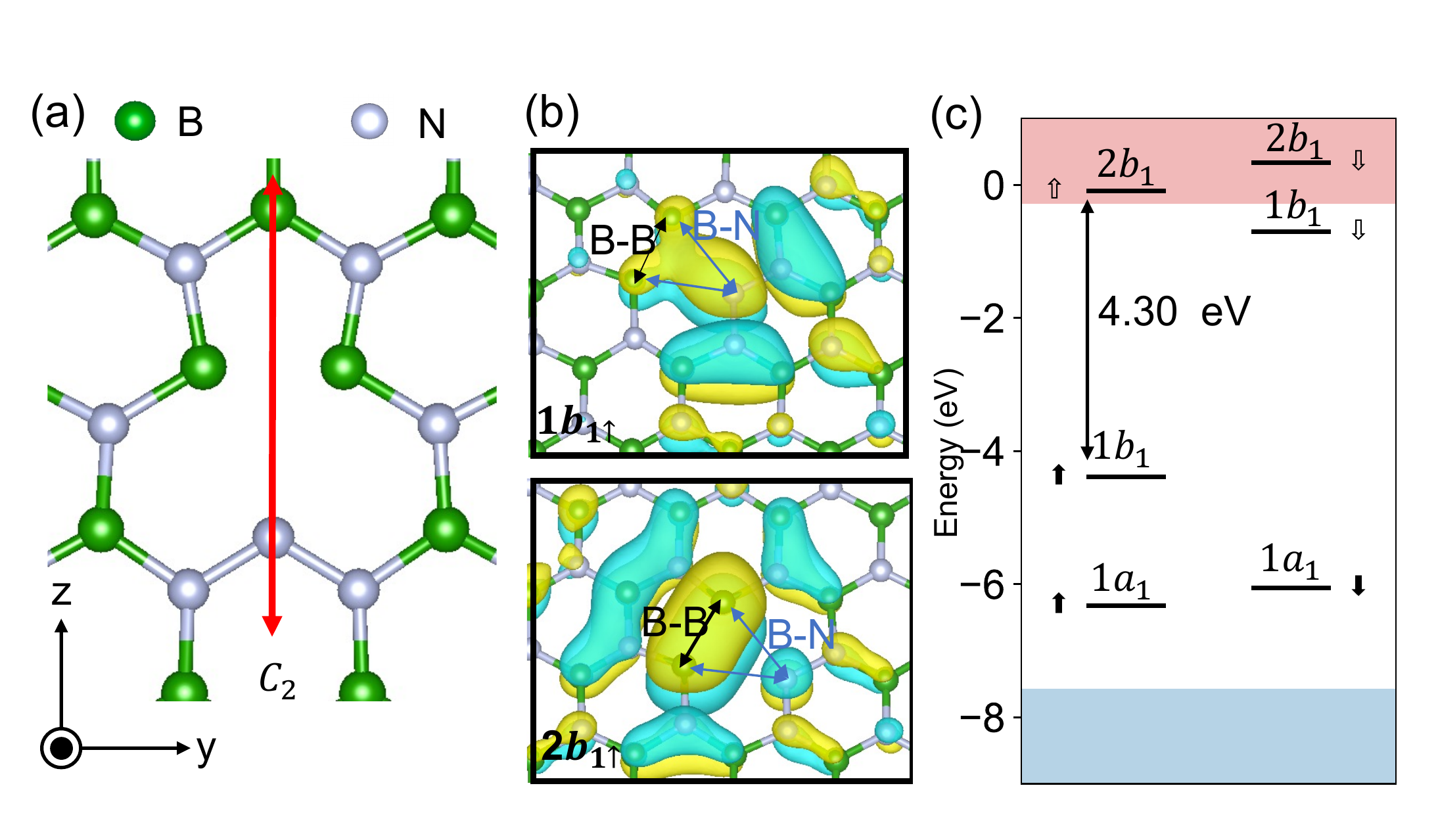}
    \caption{Structural and electronic properties of nitrogen vacancy ($\nbvn$) in hBN with $C_{2v}$ symmetry. (a) Atomic structure, (b) defect-related wave functions, and (c) single-particle diagram of ground state at $\text{G}_0\text{W}_0$@PBE level. The vacuum level and defect state notations are set up in the same way as $\cbcn$.}%\textcolor{red}{YP: move "eV" in panel c to the same line with the same font}}
    \label{Fig2}
\end{figure}

\subsection{Electronic and Optical Properties of Defects in hBN}

We start from the discussion of electronic structure and optical properties of defects in monolayer hBN at natural lattice constant, then followed by the discussion of strain and layer thickness/substrate effects. 
%effect of strain on defect properties, then followed by the discussion of layer and substrate effect. 
We choose carbon dimer ($\cbcn$) and nitrogen vacancy ($\nbvn$) defects in hBN as our prototypical systems, both of which are common defects in hBN. 
%i.e. carbon impurity as the former and vacancy defect as the latter. 
In particular $\cbcn$ was identified as a defect candidate for 4 eV SPE in hBN~\cite{mackoit-sinkevicieneCarbonDimerDefect2019}.
%The former is simple close shell substitutional defect that has been identified as a candidate for 4 eV SPE~\cite{mackoit-sinkevicieneCarbonDimerDefect2019}, and the latter one is vacancy complex which is widely considered as one of the main candidate for the 2 eV SPE~\cite{tranQuantumEmissionHexagonal2016b,jungwirthTemperatureDependenceWavelength2016,tranRobustMulticolorSingle2016b,grossoTunableHighpurityRoom2017a}.%% 
%%%%%
The atomic structures and electronic structures of $\cbcn$ and $\nbvn$ are shown in Fig.~\ref{Fig1} and Fig.~\ref{Fig2}, respectively. The atomic structure shows that both $\cbcn$ and $\nbvn$ belong to the $C_{2v}$ point symmetry group. We label the defect wave functions according to the irreducible representations to which they transform. In particular, the $1b_1$ and $2b_1$ states of $\cbcn$, and $1b_{1\uparrow}$ and $2b_{1\uparrow}$ states of $\nbvn$ are of interest, as they correspond to the optically allowed intra-defect transitions~\cite{mackoit-sinkevicieneCarbonDimerDefect2019,wuCarrierRecombinationMechanism2019}. The comparison between the electronic structures at PBE and $\gw$@PBE levels in Fig.~\ref{Fig1}c and d indicates that the quasiparticle correction shifts the occupied defect states downward and unoccupied states upward, opening up the defect gap of $\cbcn$ to 6.33 eV from 3.55 eV, and the defect gap of $\nbvn$ to 4.30 eV from 2.06 eV at PBE.\\

We then carried out BSE calculations for the related vertical excitation energy. The complete BSE spectra and the exciton wave function are presented in the SI Figs. S3 and S4. The results of $\cbcn$ indicate the presence of a single isolated peak related to the $1b_1 \rightarrow 2b_1$ transition at the BSE excitation energy ($E_{\text{BSE}}$) of 4.44 eV, with a corresponding exciton binding energy of 1.89 eV and a radiative lifetime of 1.1 ns. The zero-phonon line energy ($E_{\text{\text{ZPL}}}$) was calculated to be 4.32 eV by subtracting the Frack-Condon shift ($E_{FC}$) of 0.12 eV from the BSE excitation energy. This result is in agreement with previous studies, where a ZPL around 4.3 eV has been obtained using cDFT with hybrid functional~\cite{mackoit-sinkevicieneCarbonDimerDefect2019} or GW\&BSE with a finite-size cluster approach~\cite{winterPhotoluminescentPropertiesCarbondimer2021}

We acknowledge the intricate nature of local structural distortion and optical transition of the $\nbvn$ defect, and the related detailed discussion is presented in the SI section III. However, to keep the main text concise, we focus on the $1b_{1\uparrow}\rightarrow 2b_{1\uparrow}$ transition at $C_{2v}$ defect symmetry. The vertical transition energy for $1b_{1\uparrow}\rightarrow 2b_{1\uparrow}$ is 2.12 eV with a 57 ns radiative lifetime. The ZPL is 1.60 eV after subtracting the  $E_{FC}$ 0.52 eV from the vertical transition at BSE. Upon considering the transition to the lower symmetry ground state at $C_s$ symmetry due to out-of-plane distortion, the ZPL energy increases to 1.70 eV. 
Our result for $1b_{1\uparrow}\rightarrow 2b_{1\uparrow}$ transition related ZPL is lower than the previous calculation~\cite{abdiColorCentersHexagonal2018,liGiantShiftStrain2020} at hybrid functional, but consistent with previous BSE results at the same structure~\cite{gaoRadiativePropertiesQuantum2021}.
This highlights the important difference when excitonic effect is taken into account. 
%which suggested ZPL of 2.04 eV with PBE~\cite{wuCarrierRecombinationMechanism2019}, 3.65 eV~\cite{abdiColorCentersHexagonal2018} and 2.52 eV~\cite{liGiantShiftStrain2020} with HSE hybrid functional.

%CBCN structure
%\textcolor{del}{We first quickly demonstrate the electronic and optical signature of unstrained monolayer $\cbcn$ and $\nbvn$ systems and identify the transition that related to the emitter. Fig \ref{Fig1} show the structure and electronic properties of $\cbcn$. Such simple close shell system has been identified as one of the candidate for 4 eV SPE~\cite{mackoit-sinkevicieneCarbonDimerDefect2019}. The Single particle diagram is plotted in both PBE and GW@PBE level in 1(c) and 1(d), with the defect-associated state plotted in 1(b) and labelled by the $C_{2v}$ character table. The addition of electron exchange correlation quasiparticle correction using the GW approximation increased the defect energy gap from 3.55 eV to 6.33 eV. \\}

%______________________Strain Effect____________________________________________________
\subsection{Effect of Strain}
%\textcolor{sz}{I rewrite the whole definition part}\\
We investigate the effect of strain by applying it along two in-plane uniaxial directions, where the parallel ($\parallel$) strain denotes the strain along the $C_{2}$ axis, and the perpendicular ($\perp$) strain denotes the direction perpendicular to the $C_{2}$ axis.
The uniaxial strain here is defined as the stretching ratio of the lattice along a certain direction, with its magnitude as $\epsilon=(l-l_0)/l_0$, where $l_0$ and $l$ are the lattice lengths before and after the strain is applied. $\epsilon$ denotes the macroscopic strain on the entire system. 
%\textcolor{red}{YP: strain here is general or not local? used lattic length here}
%\textcolor{red}{Then we also define the strain tenor $\epsilon_{i,j}=\epsilon cos\theta_i cos\theta_j$~\cite{mendelsonStrainInducedModificationOptical2020},} \textcolor{sz}{where $\theta_i$($\theta_j$) is the angle between the strain and $i^{\text{th}}$($j^{\text{th}}$) coordinate direction. Due to the $C_{2v}$ symmetry, the linear static-strain } 
The strain tensor can thus be written as $\epsilon_{i,j}=\epsilon cos\theta_i cos\theta_j$~\cite{mendelsonStrainInducedModificationOptical2020} where $\theta_i$($\theta_j$) is the angle between strain axis and $i^{\text{th}}$($j^{\text{th}}$) coordinate direction (as defined in Fig.~1(a) and Fig.~2(a)).

Due to the $C_{2v}$ symmetry, the response of ZPL to the applied strain at the first order depends only on two in-plane diagonal components of the strain tensor, and thus can be written as:

\begin{equation}
    \Delta E_{\text{\text{\text{ZPL}}}} = \epsilon (\kappa_{\parallel} cos^2\theta + \kappa_{\perp} sin^2\theta), \label{eq:efiff_strain}
\end{equation}

where $\theta$ is the angle between the strain axis and $C_{2}$ axis. The energy response to strain is quantified by two linear strain susceptibilities $\kappa_{\parallel(\perp)}$. 
In order to understand the determining factors on ZPL energy change ($\Delta E_{\text{\text{\text{ZPL}}}}$) when applying strain, $\Delta E_{\text{\text{\text{ZPL}}}}$ can be decomposed into different contributions as follows.
\begin{equation}
	\Delta E_{\text{\text{\text{ZPL}}}}=\Delta E_{\text{\text{PBE}}}+\Delta E_{\text{\text{QP}}}-\Delta E_b - \Delta E_{\text{\text{FC}}}, \label{energy_sep}
\end{equation}
where $\Delta E_{\text{\text{PBE}}}$ is the change of DFT single particle energy at the PBE level due to strain; $\Delta E_{\text{\text{QP}}}$ is the change of quasipartical energy correction; $\Delta E_{b}$ is the change of exciton binding energy by solving BSE, and $\Delta E_{\text{\text{FC}}}$ is the change of excited state relaxation energy (Frack-Condon shift) under strain. 

\begin{comment}
where $E_{\text{\text{PBE}}}$ is the electronic energy difference between defect states at DFT single particle level. $\Delta E_{\text{\text{QP}}}$ is the quasiparticle energy correction from the GW approximation, which includes the many-body effect from electron correlation.
$E_{b}$ is the exciton binding energy from BSE calculations that include the electron-hole interaction. $E_{\text{\text{FC}}}$ captures the energy relaxed to the excited state's equilibrium geometry from the ground state's equilibrium geometry (after vertical excitation).
\end{comment}
%Since we find the linear response approximation also works for each term of the energies in Eq (\ref{energy_sep}) separately, the Eq (\ref{eq:efiff_strain}) can be applied to them, e.g. $E_{\text{\text{\text{PBE}}}} = \epsilon (\kappa^{\parallel}_{PBE} cos^2\theta + \kappa^{\perp}_{PBE} sin^2\theta)$.%%%
%%%%%%%%%%
We thus can separate ZPL's strain susceptibility ($\kappa_{\text{ZPL}}$) into terms corresponding to different levels of contribution from Eq.~\ref{energy_sep},
\begin{equation}
	\kappa^{\parallel(\perp)}_{\text{ZPL}}=\kappa_{\text{PBE}}^{\parallel(\perp)}+\kappa_{\text{QP}}^{\parallel(\perp)}-\kappa_{b}^{\parallel(\perp)}-\kappa_{\text{FC}}^{\parallel(\perp)}\label{eq:susceptibility}.
\end{equation}
By providing all the strain susceptibility components in Eq.~\ref{eq:susceptibility}, we can reveal the origin of ZPL response to strain.
%============== YP Revision stopped here ===================================================%

%Figure2: Strain effect results
\begin{figure*}
    \centering
    \includegraphics[width=\textwidth]{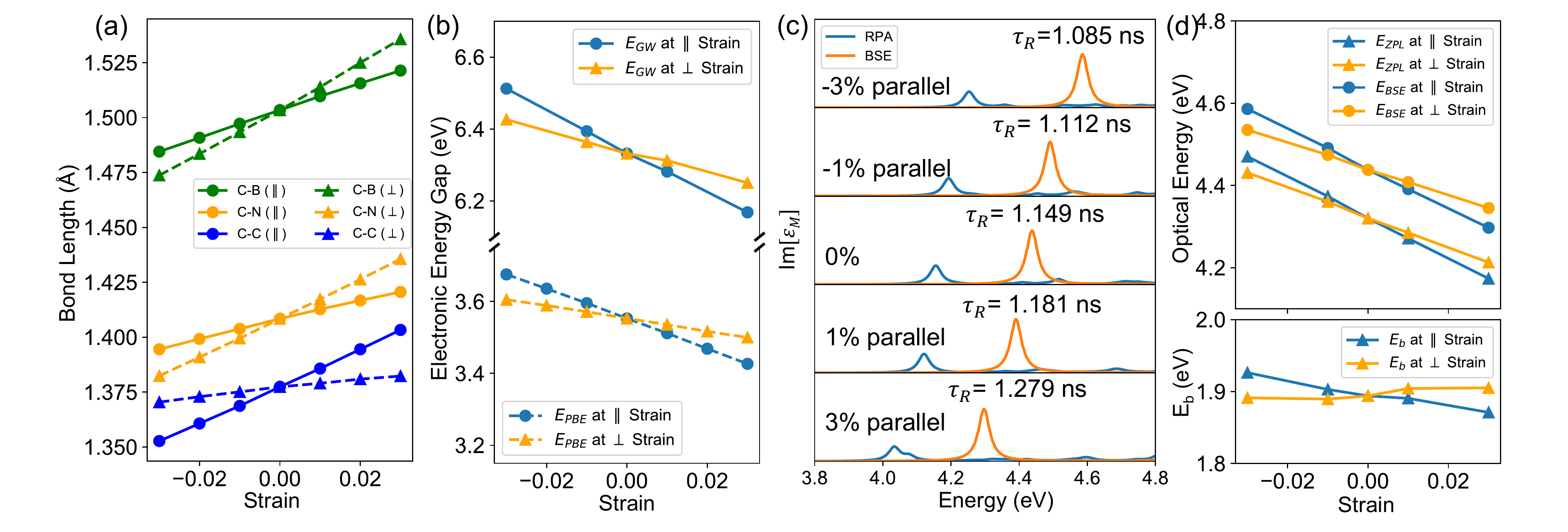}
    \caption{Strain effect on electronic and optical properties of $1b_1 \rightarrow 2b_1$ transition at $\cbcn$ in hBN.
    (a) Bond length change of the $\cbcn$ defect as parallel (to $C_{2}$ symmetry axis, $\parallel$) and perpendicular ($\perp$) strain is applied. 
(b) Electronic energy gap change between defect states, including PBE and GW@PBE 
results along two directions of strain.  (c) Optical spectra (only energy range below the bulk-state transition shown) are red shifted as the strain
increases at both RPA level (blue) and BSE level (orange). The radiative lifetime is denoted as $\tau_R$. (d) BSE excitation  energy and ZPL energy (up), exciton binding energy ($E_b$, down).%\textcolor{red}{YP: one figure can't use the same symbol for two different quantities- change GW ones}
}
    \label{Fig3}
\end{figure*}

%Plot 2 
The effect of strain on the structural, electronic, and optical properties of $\cbcn$ in monolayer hBN is described in Fig.~\ref{Fig3}. Fig.~\ref{Fig3}(a) displays the change in three bond lengths, including the defect-defect bond length (C-C) and the defect-nearest-neighbor bond lengths (C-N and C-B). All three bond lengths increase linearly with strain (where a positive sign denotes stretching strain and a negative sign denotes compressing strain). Fig.~3(b) illustrates the change in defect electronic energy gap at PBE level ($E_{\text{\text{PBE}}}$) and G$_0$W$_0$@PBE level ($E_{\text{\text{GW}}}$). Fig.~\ref{Fig3}(c) and (d) present the optical properties of the defect emitter, including the absorption peak related to $1b_1 \rightarrow 2b_1$ transition at RPA and BSE levels (c), and the corresponding BSE excitation energy ($E_{BSE}$), ZPL ($E_{ZPL}$), and exciton binding energy ($E_b$) as a function of strain (d). Both the optical and electronic energy gaps of the defect-related transition exhibit a linear red shift with increasing strain, with the parallel strain resulting in a larger response compared to the perpendicular strain. However, the radiative lifetime ($\tau_R$) and exciton binding energy ($E_b$) show negligible response to the strain.

%\textcolor{red}{the above paragraph needs to be more concise and better clarity; it reads like a laundry list, some reorganization of sentences are needed here. }
%Table discussion
%CBCN table
\begin{table*}
\caption{\label{tab1} The bond length change rate ($R$) and strain susceptibilities ($\kappa$, with unit of meV$/\%$) of $\cbcn$ and $\nbvn$ under the parallel ($\parallel$) and perpendicular ($\perp$) strain. The subscript of $R$ denotes the atomic distance/chemical bond of interest, surrounding the defect center. The subscript of $\kappa$ represents various contributions, i.e. ``QP" denotes quasiparticle correction; ``b" denotes exciton binding energy; ``FC" denotes Frank condon shift; ``ZPL" denotes zero-phonon lines.  %\textcolor{red}{YP: digits need to be consistent, e.g. all $\kappa$ to the second digit. SZ: it is to the second digits}
}
\begin{ruledtabular}
\begin{tabular}{ccccccccc}
 \multicolumn{9}{c}{$\cbcn$}\\
 \hline
 &$R_{\text{C-C}}$&$R_{\text{C-B}}$&$R_{\text{C-N}}$& $\kappa_{\text{PBE}}$
  %\footnote{The unit of $\kappa_{\text{PBE}}$,$\kappa_{\text{QP}}$,$\kappa_{\text{GW}}$,$\kappa_{b}$,$\kappa_{\text{FC}}$,$\kappa_{\text{ZPL}}$ is meV/\%} 
 &$\kappa_{\text{QP}}$ &$\kappa_{b}$& $\kappa_{\text{FC}}$& $\kappa_{\text{ZPL}}$\\
\hline
$\parallel$ &0.613&0.301&0.411&-41.58&-15.70&-8.90&1.30&-49.63\\
$\perp$ &0.144&0.629&0.688&-17.75&-11.40&2.82&4.64&-36.49\\
\hline
 \multicolumn{9}{c}{$\nbvn$}\\
 \hline
 &$R_{\text{B-B}}$&$R_{\text{B-N}}$& - & $\kappa_{\text{PBE}}$ &$\kappa_{\text{QP}}$ & $\kappa_{b}$ & $\kappa_{\text{FC}}$& $\kappa_{\text{ZPL}}$\\
\hline
$\parallel$ &0.463&1.664& - &-59.70&-19.32 & -14.27 & 40.35 &-105.20\\
$\perp$ &2.072&0.490& - &44.96& 1.00 & 2.05 & 6.31 & 38.02\\
\end{tabular}
\end{ruledtabular}
\end{table*}

%\textcolor{sz}{rewrote, add logic}\\
%Strain susceptibility discussion
The linearity of response to strain in Fig.~\ref{Fig3} suggests that the linear response model represented by Eq.~\ref{eq:efiff_strain} and Eq.~\ref{eq:susceptibility} is adequate. As a result, the strain susceptibility and bond length change rate ($R_{\nu}$; related to the local bond length change speed under strain, will be defined later) are summarized in TABLE~\ref{tab1}. Similar calculations were performed for the $\nbvn$ defect system, summarized in the same table as well.
%%%%We listed the bond length change rate ($R_{\nu}$) of several defect related bonds and strain susceptibilities component from each levels of contribution.\\

The results of ZPL strain susceptibility ($\kappa_{ZPL}$) reveal that the $\cbcn$ defect exhibits a similar negative strain susceptibility (red shift of ZPL) in both $\parallel$ and $\perp$ components, i.e. $\kappa^{\parallel}_{\text{ZPL}} = -49.63$ meV/\% and $\kappa^{\perp}_{\text{ZPL}} = -36.49$ meV/\%. On the other hand, the $\nbvn$ defect exhibits a disparate strain response behavior in two directions, where the $\parallel$ component of strain susceptibility is negative ($\kappa_{\text{ZPL}}^{\parallel} = -105.20$ meV/\%) and the $\perp$ component is positive ($\kappa_{\text{ZPL}}^{\perp} = 38.02$ meV/\%). The sign difference on strain susceptibility reflects the different bonding nature of defects as discussed later. By substituting the two components of strain susceptibilities into Eq. \ref{eq:efiff_strain}, one can determine the ZPL energy shift under any uniaxial strain in the linear response regime.
%%%%
Previous experimental work has shown that the uniaxial strain susceptibility of 2 eV SPE ranges from -120 meV/\% to 60 meV/\% without specifying the strain direction~\cite{mendelsonStrainInducedModificationOptical2020,hayeeRevealingMultipleClasses2020,grossoTunableHighpurityRoom2017a}. 
Therefore, our calculated strain response for the $\nbvn$ defect falls within the experimental strain susceptibility range~\cite{xueAnomalousPressureCharacteristics2018}(more related discussion detailed in SI). 

Our analysis of $\kappa_{\text{ZPL}}$ composition in Table~\ref{tab1} indicates that the determining factors of strain response is different between two defect systems. 
For $\cbcn$ defect, the change from single-particle level at PBE and GW ($\kappa_{\text{PBE}}$ and $\kappa_{\text{QP}}$) has the dominant contribution, while exciton binding energy $\kappa_{b}$ and the Frank-Condon shift $\kappa_{\text{FC}}$ have a negligible impact.
%
%is attributed to the single-particle energy level difference of two defect states, namely $\kappa_{\text{PBE}}$ and $\kappa_{\text{QP}}$. The contribution from exciton binding energy can change the disparities between the parallel and perpendicular component of the strain susceptibility, and thus affecting angular dependence of optical strain response. On the other hand, the Frank-Condon shift is found to have a negligible impact. 
%
On the other hand, for the $\nbvn$ defect, although $\kappa_{\text{PBE}}$  and $\kappa_{\text{QP}}$ still dominate, the other contributions from 
$\kappa_{b}$ and $\kappa_{\text{FC}}$
have a sizable impact for $\kappa^{\parallel}_{\text{ZPL}}$ (not for $\kappa^{\perp}_{\text{ZPL}}$).
%This distinct strain response behavior between two defects may be understood by molecular orbital theory as explained later.  
%
%
%The contribution from exciton binding energy, Frank-Condon shift, and the quasiparticle energy have a non-negligible impact on the parallel strain susceptibility, whereas the perpendicular strain response remains insensitive to these factors. This distinguished strain response behaviour along two direction indicate that they should derived from different nature of molecular orbital. 
%
%
Our analysis of the strain susceptibility highlights the importance of many-body effects and excited-state relaxation in determining the optical strain response. These factors impact both the magnitude and anisotropicity of the response.
In light of these findings, it is crucial to consider these effects in the study of strain engineering for optical spectroscopy.

We then discuss the bond length change rate ($R_{\nu}$) of local atomic distance %defects
$\nu$ (TABLE \ref{tab1}) to identify the most relevant molecule orbitals responding to strain. The bond length change rate is defined as $(d_{\nu}-d_{0\nu})/(d_{0\nu})$, with $d_{0\nu}$ and $d_{\nu}$ as the local atomic distances before and after applying the strain. This quantity indicates to what extent the macroscopic tensile/compression strain can be transferred into the microscopic local structural change. This helps us develop insights on optoelectronic properties based on molecular orbital theory, given the localized nature of defect-related wave functions.

For example, in the $\cbcn$ defect system, C-C bond has the largest change under the parallel ($\parallel$) strain to $C_{2}$ axis (therefore the largest $R_{C-C}$), which induces change to the corresponding molecular orbitals (MO) between two carbon atoms. From the defect wavefunctions in Fig.~\ref{Fig1}b, we can find the lower defect level $1b_1$ has a $\pi$ bonding character between two C atoms; instead, the higher defect level $2b_1$ has a $\pi^*$ antibonding character. As a result, the stretching of the C-C bond weakens charge density overlap between two C atoms, leading to a decrease of energy gap between  $\pi$ ($1b_1$) and $\pi^*$ ($2b_1$), shown as a red shift in Fig.\ref{Fig3}(b) and (c). This change also results in a negative strain susceptibility in Table I. Similar discussions can be applied to the $\nbvn$ defect system as well; one exception is that the strain susceptibility is positive when applying strain perpendicular to the $C_{2}$ axis of $\nbvn$ defect, where we found the B-B distance has the largest change ($R_{B-B}$ close to 1). Interestingly, the highest occupied defect level $1b_1$ has a nearly non-bonding character between two B atoms, but the lowest unoccupied defect level $2b_1$ in the same spin channel has a bonding character between two B atoms. Stretching B-B bond will decrease the charge density overlap between two B, which increases the energy of $2b_1$ but weakly affects $1b_1$. As a result, the energy gap between two defect levels is increased, which explains its positive $\kappa$ along $\perp$ direction, opposite to the others in Table~\ref{tab1}.

In summary, our study analyzed the effects of strain on the electron and optical properties of hBN defects through the use of two representative systems: the carbon-dimer defect $\cbcn$ and the nitrogen vacancy complex $\nbvn$. Our calculations included both many-body effects and relaxation of excited states in determining the strain susceptibility of ZPL. Our findings emphasized the importance of incorporating many-body contributions in studies of optical spectroscopy under strain. Additionally, we analyzed the different signs of strain response susceptibility through molecular orbital theory, after identifying the primary molecular orbitals responding to strain.\\

 %______________________Layer and Substrate____________________________________________________

\subsection{Layer Thickness Dependence and Substrate Effect}

%Figure3: layer and substrate structure
\begin{figure}
    \centering
    \includegraphics[width=\linewidth]{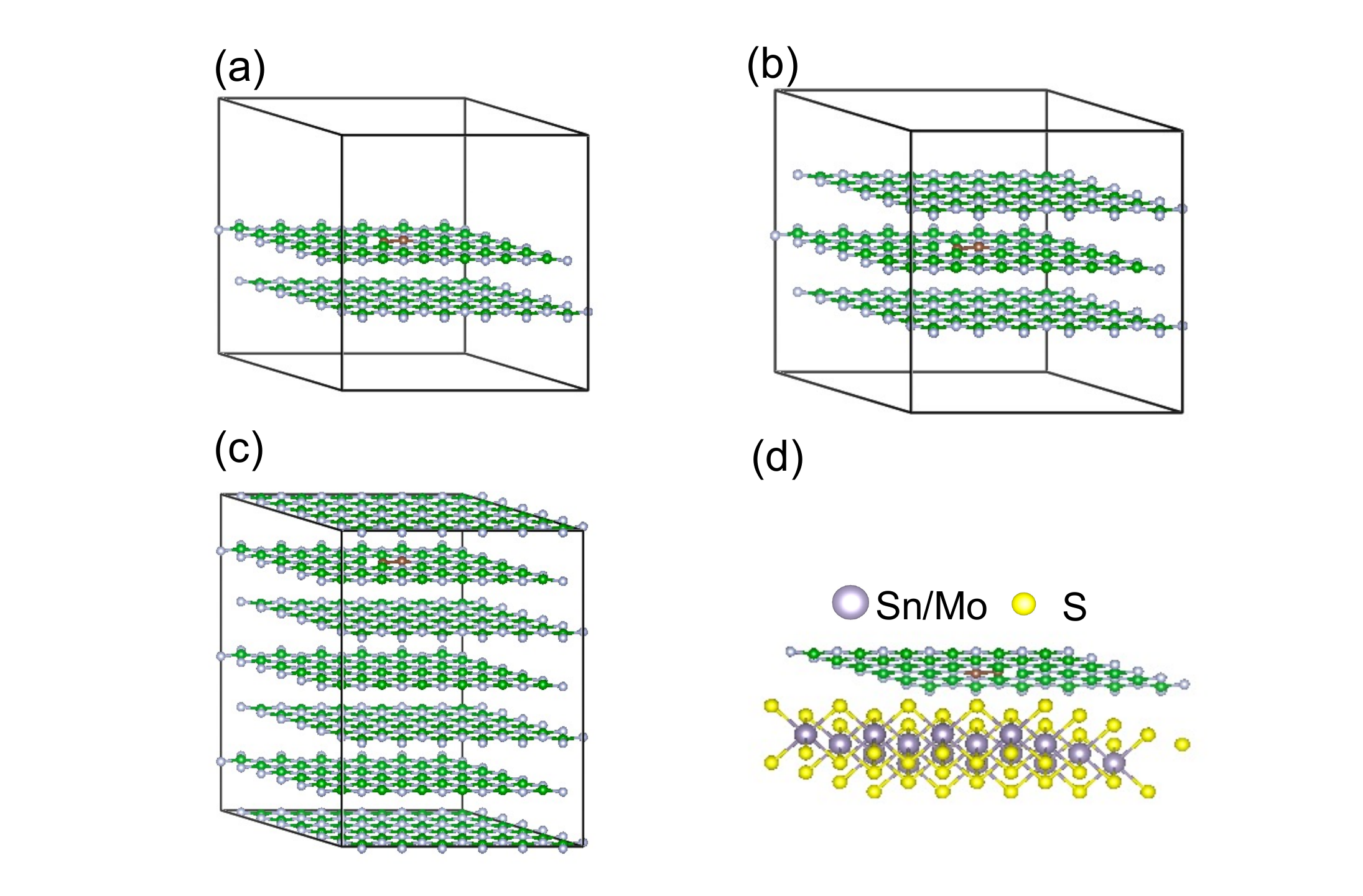}
    \caption{Lattice structures of (a) defect in double-layer hBN, (b) defect in three-layer hBN, (c) defect in bulk hBN, (d) defect in monolayer hBN on $\mos2$ and $\sns2$ substrates.}
    \label{Fig4}
\end{figure}

We next look at the layer thickness dependence and substrate effects on defect emitter properties. We use our implicit $\chi_{\text{eff}}$-sum method~\cite{guoSubstrateScreeningApproach2020,guoSubstrateEffectExcitonic2021} to calculate properties of one isolated defect within 1-3  layers (Fig.\ref{Fig4}(a-b)) and bulk hBN (Fig.\ref{Fig4}(c)).
We choose the AA' stacking structure and set the inter-layer distance to the bulk value of of 3.33 \AA{}~\cite{hodGraphiteHexagonalBoronNitride2012}. 
%\textcolor{red}{YP: did you relax it? one should use relaxed lattice constant in principles.} %it just means, we choose the number from the citation. in citation this value should come from both experiment calculation, but we don't need to explain more here.
%\textcolor{red}{(based on what?)} \textcolor{kc}{this expt. paper gives lattice constant of hBN (Appl. Phys. A 75, 431–435 (2002))}\textcolor{sz}{SZ: reference added}. 
For the substrate effect study,  we choose two different transition metal dichalcogenides (TMD) materials as substrates (Fig.\ref{Fig4}(d)) with layer 
distance to defective hBN 3.31 \AA{} for $\sns2$ and 3.33 \AA{} for 
$\mos2$ substrates~\cite{guoSubstrateScreeningApproach2020}. % 

\begin{figure}
\centering
    \includegraphics[width=\linewidth]{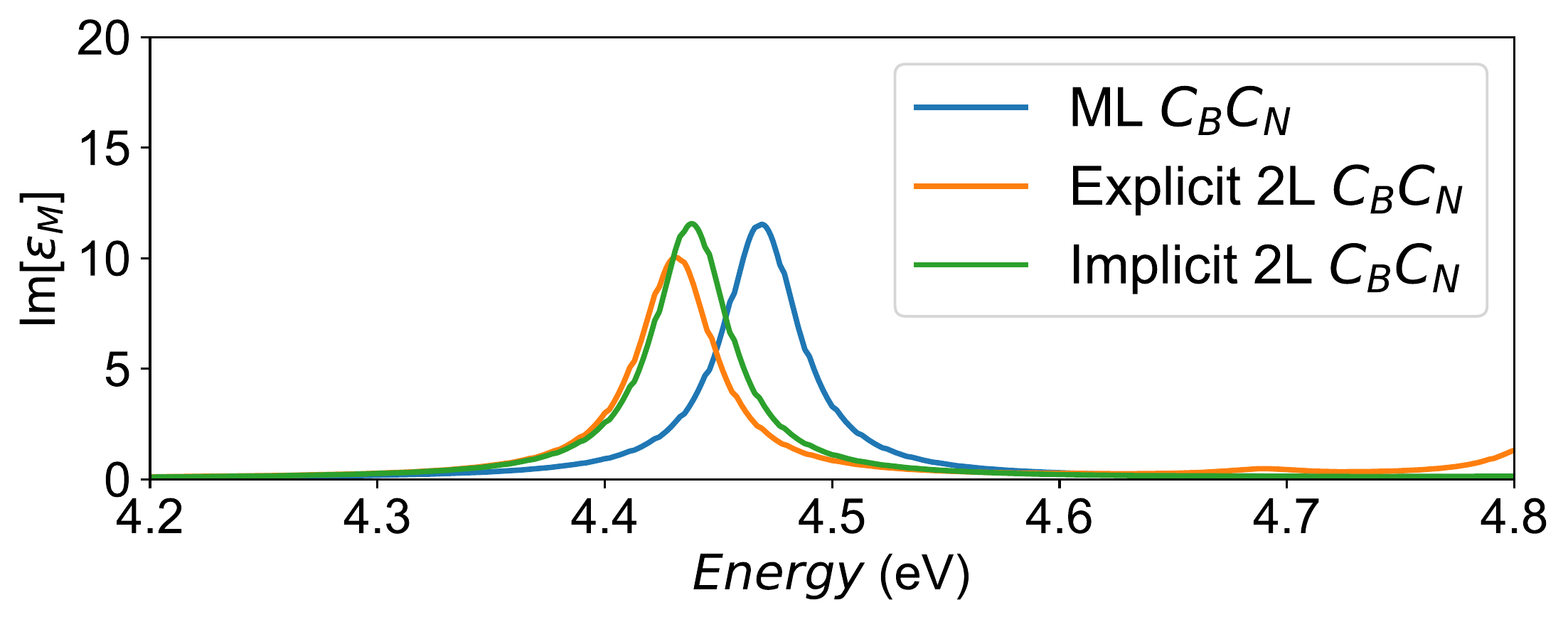}
    \begin{ruledtabular}
    \begin{tabular}{ccccc} 
                     &$E_{\text{BSE}}$(eV) & $\tau_{R}$ (ns) &$E_{\text{GW}}$(eV)& $E_{b}$(eV) \\           
     \hline
     ML& 4.469&1.1 &6.336 & 1.867\\
    Implicit &   4.438  &        1.3        &          6.045            &   1.607 \\
    Explicit &   4.431  &        1.3        &          5.958            &   1.527 \\
    Error(Im.-Ex.) &0.007&       $<0.01$       &          0.087            &   0.080 \\
    \end{tabular}
    \end{ruledtabular}
    \caption{
    Benchmark calculations of $\cbcn$ in two-layer hBN for $\chi_{eff}$-sum implicit method. The 
    figure above shows the absorption spectra at BSE. 
    The table below listed the BSE defect peak transition energy ($E_{\text{BSE}}$), 
    radiative lifetime ($\tau_{R}$), electronic energy gap between defect states at GW@PBE 
    level ($E_{\text{\text{GW}}}$), and exciton binding energy of the defect peak ($E_b$). 
     The errors (Error) are the difference of results obtained by implicit and explicit methods. We noticed there is an error cancellation between $E_{\text{GW}}$ and $E_b$.}%\textcolor{red}{The error of GW is larger than the shift in spectra. } }
    \label{Fig5}
\end{figure}

To validate the result from our implicit $\chi_{\text{eff}}$-sum method~\cite{guoSubstrateScreeningApproach2020,guoSubstrateEffectExcitonic2021}, we compare its results with explicit bilayer calculations by using $\cbcn$ in hBN as an example. 
%
%we use bilayer $\cbcn$ as a benchmark, where the explicit result is obtained by direct two layer atomic structure calculation and implicit result is from $\chi_{\text{eff}}$-sum method calculation with single layer. 
In Fig.\ref{Fig5}, the panel above is the BSE absorption spectrum of defect-related peaks, where we show the explicit (orange) and implicit (green) two-layer calculations give similar results for carbon-dimer defect, which are both red-shifted by 0.03 eV compared to the monolayer one (blue). The table below summarizes the excitation energy ($E_{\text{BSE}}$), electronic gap ($E_{\text{GW}}$), and radiative lifetime ($\tau_{R}$)
of the defect emitter, calculated from explicit and implicit interface methods. 
%Fig.\ref{Fig5} shows the error of implicit $\chi_{\text{eff}}$-sum method~\cite{guoSubstrateScreeningApproach2020,guoSubstrateEffectExcitonic2021} for optical energy is smaller than 10 meV, comparable to the error bar of GW/BSE calculations, negligible in our discussions. 

\begin{comment}
\textcolor{del}{The electronic energy gap error is up to 0.087 due to the inter-layer wave function hybridization.} \textcolor{magenta}{that error does not come from hybridization only}
\textcolor{del}{However there exist an error cancelling between the binding energy and electronic energy, so as a result the implicit method is accurate with the error smaller than 0.01 eV for both optical energy and radiative lifetime. }\textcolor{magenta}{I suggest not to discuss such ``error cancellation" staff in the main text, it's very misleading}
\end{comment}
%Figure5: Benchmark spectrum

%______________________Layer____________________________________________________
%Description for BSE peak, layer effect
%Figure6:
\begin{figure*}
    \centering
    \includegraphics[width=\textwidth]{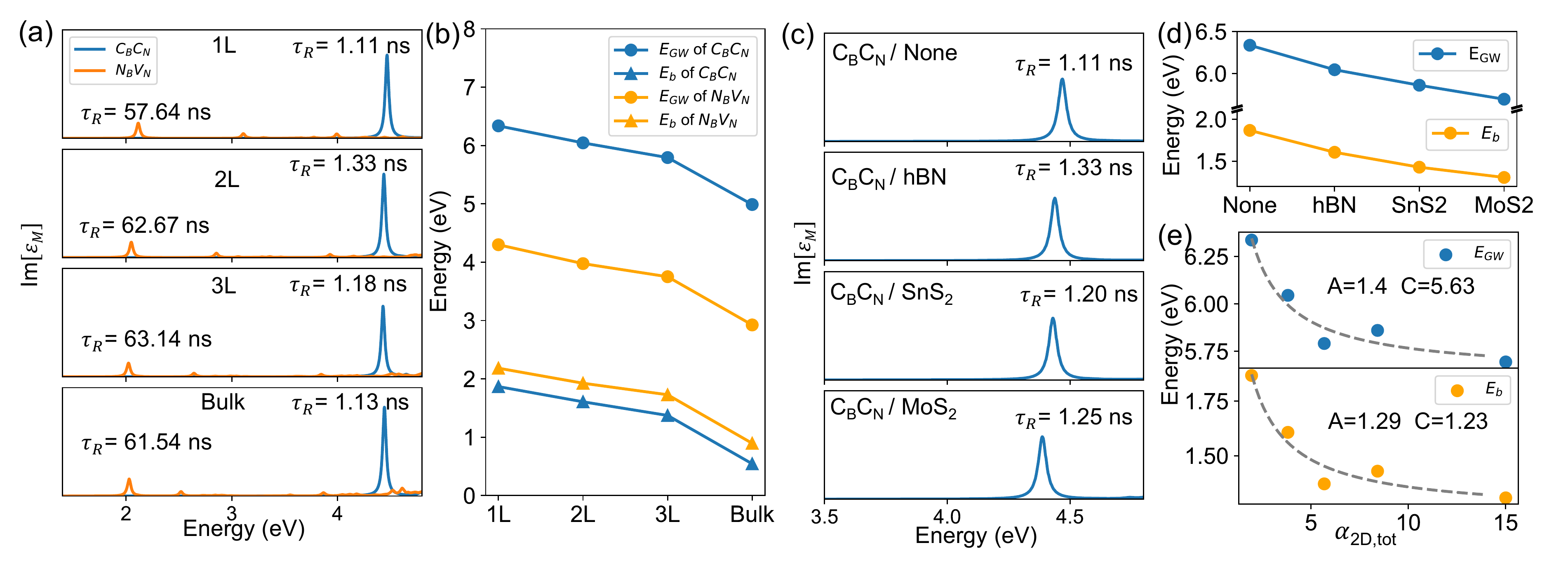}
    \caption{Layer thickness and substrate effect on electronic and optical properties. 
    (a)The BSE peak of $\cbcn$ and $\nbvn$ defects with different layer number from 1L to bulk. (b) The exciton binding energy and GW energy of $\cbcn$ and $\nbvn$ defects as a function of layer thickness. (c) The BSE optical spectra of $\cbcn$ defect in monolayer hBN on various substrate. (d) The exciton binding energy and GW energy gap of $\cbcn$ defect in monolayer hBN on different substrates. 
    (e) The GW energy and exciton binding energy as the function of total 2D polarizability ($\alpha_{2D, tot}$). The grey dashed line is the fitting result. }
    \label{Fig6}
\end{figure*}
We then show the results of layer thickness and substrate effects on optical spectra by using implicit $\chi_{\text{eff}}$-sum method in Fig.\ref{Fig6} (a) and (c). With increasing the number of layers or adding substrates, the position of defect peaks remains nearly constant, with tiny shift (within 80 meV).  
The radiative lifetime ($\tau_R$) also has negligible change. This result is consistent with experimental observations, where defect emitters did not change its ZPL energy either from 1L to 5L or with various substrates~\cite{krecmarovaExtrinsicEffectsOptical2021}.\\\indent
In Fig.\ref{Fig6}(b) and (d) we show the layer thickness and substrate effect on GW energy gap ($E_{\text{GW}}$) and exciton binding energy($E_b$) of intra-defect transitions. A monotonic  decrease of both $E_{\text{GW}}$ and $E_b$ has been observed as increasing layer thickness, where the change nearly cancels each other, leaving the optical excitation energy unchanged ($E_{\text{opt}}=E_{\text{GW}}-E_b$).
Interestingly, the substrate/layer effects on two defect emitters are similar; e.g. for $\cbcn$, $E_{\text{GW}}$ decreases by 1.348 eV from monolayer to bulk, and for $\nbvn$ it  decreases by 1.370 eV.
%$E_{\text{GW}}$ and $E_b$ energy renormalization of two defect emitters are similar. The $E_{\text{GW}}$ for $\cbcn$ emitter decreased by 1.348 eV from monolayer to bulk, and for $\nbvn$ decreased by 1.37 eV. 
%
This finding indicates that the defect state energy renormalization by layer thickness and substrates weakly depends on the specific defects, but is mostly determined by the host environment.
%______________________2D polarizability______________________
%\textcolor{red}{YP: sentence above needs rewritten}\textcolor{sz}{SZ: paragraph rewrote}\\

We then qualitatively show that the energy renormalization of $E_{\text{GW}}$ and $E_{b}$ of defects is directly related to the environmental dielectric screening surrounding the defects, represented by total 2D effective polarizability ($\alpha_{2D,tot}$)~\cite{molina-sanchezMagnetoopticalResponseChromium2020,tianElectronicPolarizabilityFundamental2020a}. $\alpha_{2D,tot}$ can be calculated by summing up the 2D effective polarizability from each subsystem ($\alpha_{2\text{D},tot}=\alpha_{2\text{D},def}+\alpha_{2\text{D},sub}$; ``sub" denotes substrate and ``def" denotes defected hBN monolayer), where the 2D effective polarizability of the subsystem is obtained by $\alpha_{2\text{D}}=(\epsilon-1)\cdot L/(4\pi)$. Here $L$ is the supercell lattice constant along the out-of-plane direction, and $\epsilon_{sub}$ is the in-plane component of macroscopic dielectric tensor of the substrate calculated by Density Functional Perturbation Theory (DFPT)~\cite{baroniPhononsRelatedCrystal2001a} (The dielectric constants are listed in SI).
We then use equation $E=A/\alpha_{2\text{D},tot}+C$ to fit the relation between  $\alpha_{2D,tot}$ and $E_{\text{GW}}$/$E_{b}$, with $A$ and $C$ the fitting parameters. The results are plotted in 
 Fig.\ref{Fig6}(e), which shows the inversely proportional relation 
can well describe the $E_{\text{GW}}$ and $E_{\text{b}}$ dependence on environmental screening surrounding the defects (from both host hBN and substrates).

To better understand the insensitivity of defect optical transition towards environmental screening, we present an analytic model 
for analyzing the renormalization of the defects' exciton binding energy ($E_{b}$). Our analysis suggests that the stability of the defect spectroscopic peak position against the layer thickness and substrate is a result of two factors: (i) the high localization of the defect wave function and (ii) the defect-related optical transition with no mixing with other transition involving delocalized wave function from the host, as discussed in detail below.
%With the above conditions the renormalization of $E_{\text{GW}}$ and $E_{b}$ should less depend on specific system and largely cancel each other, causing the rubustness of the optical energy.
%Model derivation:

We start with the Hamiltonian for Bethe-Salpeter equation in Wannier basis \cite{bechstedtManyBodyApproachElectronic2015}:
\begin{eqnarray}
H^S(cv,c'v',\Vec{R}) = E_{\text{\text{GW}}}\delta_{cc'}\delta_{vv'} 
- W_{vv'}^{cc'}(\Vec{R}) &\nonumber \\
+ 2\delta_{S0} \sum_{\Vec{R'}} \Bar{V}_{v'c'}^{cv}(\Vec{R'})\delta_{\Vec{R}0}\label{eq:bse}&
\end{eqnarray}

where $v/c$ are the pairs of occupied and unoccupied states, $\Vec{R}$ is the real-space lattice site, $S$ is the spin, $W$ is the screened exchange interaction between electron and hole (the direct term), $\Bar{V}$ is the exchange term from the Hartree potential,   and $E_{\text{\text{GW}}}$ is the GW electronic energy gap between $c$ and $v$ states. The Wannier function basis is obtained from the Fourier Transform of the reciprocal-space Kohn Sham wave function to the real-space lattice :
\begin{equation}
    a_{v/c}(\Vec{x}-\Vec{R})=\frac{1}{G^{3/2}}\sum_k e^{-i\Vec{k}\cdot \Vec{R}}\phi_{v/c,k}(\Vec{x})\label{eq:wannier}
\end{equation}

The exciton energy and wavefunctions are obtained by diagonalizing the Hamiltonian:
\begin{equation}
    \sum_{c'v'} H^s(cv,c'v')\phi_\lambda(c'v') = E_\lambda\phi_\lambda(cv)
\end{equation}

where $\phi_\lambda$ is the of exciton wavefunction and $E_\lambda$ is the exciton energy. 

We consider the perturbation from the dielectric screening of pristine substrates or other pristine host material layers. (Notice that the in-homogeneous local dielectric environment from the host material around the defect is part of the unperturbed Hamiltonian.) The first order perturbed change to the exciton energy $E_\lambda$ is then:
\begin{eqnarray}
    \Delta E_\lambda &&= \bra{\phi_\lambda}\delta \hat{H^s} \ket{\phi_\lambda} \nonumber \\ 
    &&=\sum_{cv}\sum_{c'v'}\braket{\phi_\lambda}{cv}\bra{cv}\delta \hat{H^s} \ket{c'v'}\braket{c'v'}{\phi_\lambda}
    \label{deltaH}
\end{eqnarray}
where the matrix element of perturbed Hamiltonian is:
\begin{eqnarray}
\bra{cv}\delta \hat{H^s} \ket{c'v'}&=&\delta H^S(c,v,c',v')\nonumber \\
&=&\delta E_{\text{\text{GW}}}^{c,v}\delta_{c,c'}\delta_{v,v'}-\delta W_{vv'}^{cc'} \label{H_matrix_element}
\end{eqnarray}
in which the first term is the renormalization of electronic gap between the pair of c/v states. The second term is the change of screened exchange interaction, which contributes to the exciton binding energy change $\delta E_b$, defined as follows:

\begin{eqnarray}
    \delta E_b &&= \bra{\phi_\lambda}\delta W \ket{\phi_\lambda} \nonumber \\ 
    &&=\sum_{cv}\sum_{c'v'}\braket{\phi_\lambda}{cv}\bra{cv}\delta W\ket{c'v'}\braket{c'v'}{\phi_\lambda}.
    \label{deltaEb}
\end{eqnarray}

Since the exchange term from the Hartree potential $\Bar{V}$ does not depend on dielectric screening, it does not appear in the perturbative Hamiltonian.
%\textcolor{red}{YP: did you write out $\delta E_b = ??$ anywhere?}

Here we consider defect related Frenkel-like excitons, whose wavefunctions usully have components only from defect states, i.e the $\braket{\phi_\lambda}{cv}$ is nonzero only when c and v are defect states.
The previous study on the GW band gap renormalization effect of dielectric screening on benzene systems~\cite{neatonRenormalizationMolecularElectronic2006}(with COHSEX model) has suggest that, when $\delta W (\Vec{x}, \Vec{x}')$ is smooth and slowly varying over the spacial extension of the orbital, the GW energy gap renormalization can be approximated by $\delta E_{GW}^{c,v}=P_{v}+P_{c}$ , where $P_{c/v}$ is the static polarization integral for c/v state,
\begin{eqnarray}
    P_{c/v} = \frac{1}{2}\int d\Vec{x}\int d\Vec{x'}a_{c/v}(\Vec{x})a_{c/v}^*(\Vec{x}')& \nonumber \\
    \delta W(\Vec{x},\Vec{x'})
    a_{c/v}(\Vec{x}')a_{c/v}^*(\Vec{x}).&
\end{eqnarray}
When defect orbitals are highly localized in real space, the integration can be reduce to the classical limit, where $P_{c/v}=\text{lim}_{\rho\rightarrow 0}\delta W(\rho)$~\cite{choEnvironmentallySensitiveTheory2018,neatonRenormalizationMolecularElectronic2006}. 

The same argument can be applied to the second term in Eq.~\ref{deltaH}, where the matrix element of two particle screened Coulomb interaction change can be written as 
\begin{eqnarray}
    \delta W_{vv'}^{cc'}(\Vec{R})=\int_{U.C.} d\Vec{x}\int_{U.C.} d\Vec{x'} 
    a_{c}^*(\Vec{x})a_{c'}(\Vec{x})&  \nonumber \\
     \delta W(\Vec{x}+\Vec{R},\Vec{x'})a_{v}(\Vec{x'})a_{v'}^*(\Vec{x'})&.
\end{eqnarray}
Since a single defect is not subject to a spacial periodicity, we can ignore the inter-site interaction and set $\Vec{R}$=0. Then we rewrite the equation with 
two-electrons' distance $\rho=|x-x'|$, assuming the substrate screening being homogeneous in-plane (which is a reasonable approximation according to our past work~\cite{guoSubstrateScreeningApproach2020}), 
%the change of screening exchange interaction only depends on the two-electrons' distance $\rho=|x-x'|$, %due to the homogeneity of the substrate screening.

\begin{eqnarray}
    \delta W_{vv'}^{cc'}=\int_{U.C.} d\Vec{x}\int_{U.C.} d\Vec{x}' 
    a_{c}^*(\Vec{x})a_{c'}(\Vec{x})& \nonumber \\
     \delta W(\rho)a_{v}(\Vec{x}')a_{v'}^*(\Vec{x}').
\end{eqnarray}

Given localized (Frenkel) exciton, $\rho$ is rather small or exciton wavefunction is spatially confined. With orthonormal condition of single-particle wavefunctions, we have 
%the same simplification as first term can be applied. 
\begin{eqnarray}
\delta W_{vv'}^{cc'}&\approx&\text{lim}_{\rho\rightarrow 0}\delta W(\rho)\int_{U.C.} d\Vec{x}\int_{U.C.} d\Vec{x}'  \nonumber \\
    &&a_{c}^*(\Vec{x})a_{c'}(\Vec{x}) 
    a_{v}(\Vec{x}')a_{v'}^*(\Vec{x}') \nonumber \\
    &=& \text{lim}_{\rho\rightarrow 0}\delta W(\rho)\delta_{c,c'}\delta_{v,v'}\label{eq:limrho}
\end{eqnarray}
Combining the discussions above on both terms in Eq.\ref{H_matrix_element}, we have the $\bra{cv}\delta \hat{H^s} \ket{c'v'}=0$ when $c$ and $v$ are both defect single-particle states. Thus, the Frenkel-like exciton,  composed by localized defect-defect transitions, will experience no change at presence of substrate screening.

At last, we apply a static model for $\delta W(\Vec{x},\Vec{x}')$ derived from image charges~\cite{choEnvironmentallySensitiveTheory2018} and compare with \textit{ab-initio} GW/BSE results. The environmental screening potential can be modeled as the potential energy of two charges in the defect layer (with dielectric constant $\epsilon_{def}$) on top of substrate (with dielectric constant $\epsilon_{sub}$)~\cite{choEnvironmentallySensitiveTheory2018}. We assume the region above the defect layer to be a vacuum. 
\begin{eqnarray}
&& W(\Vec{x},\Vec{x}')=W(\rho) = \frac{1}{\epsilon_{def} \rho} + 2\sum_{n=1}^{\infty}\frac{L_{12}^n}{\epsilon \{\rho^2 + (2nd)^2)\}^{1/2}} \nonumber \\
&&+L_{12}\sum_{n=0}^{\infty}\frac{ L_{12}^n}{\epsilon (\rho^2 + \{[(2n+1)d]^2)\}^{1/2}}
\label{Wmodel}
\end{eqnarray}
where $d$ is the thickness of the defect layer, and $L_{12}=(\epsilon_{def}-\epsilon_{sub})/(\epsilon_{def}+\epsilon_{sub})$.
We use the monolayer defect system without any substrate ($\epsilon_{sub}=1$) as the reference, then $\delta W$ is defined as the change of $W$ with and without substrates:
%for the screened potential change $\delta W$ under the effect of substrates or host material layers. 
\begin{equation}
	\delta W(\rho,\epsilon_{sub}) = W(\rho,\epsilon_{sub}) - W(\rho, \epsilon_{sub}=1). \label{DeltaW}
\end{equation}
\begin{figure}[h]
    \centering
    \includegraphics[width=\linewidth]{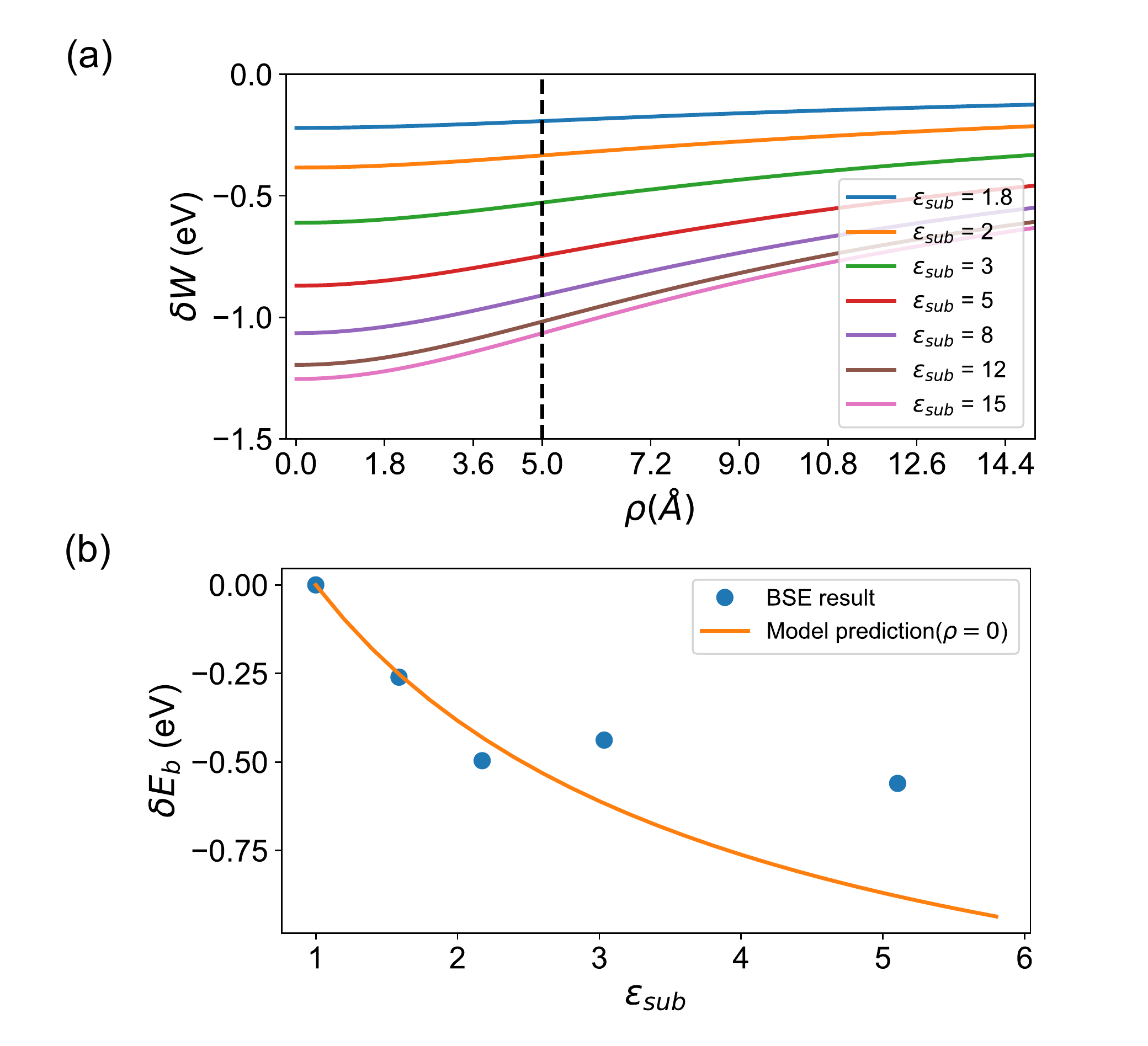}
    \caption{(a) The change of screened Coulomb potential as a function of electron-hole separation ($\rho$) at different substrate dielectric constants $\epsilon_{sub}$. The calculation is performed using Eq.~\ref{Wmodel} and Eq.~\ref{DeltaW}. The dash line shows the localization range of defect wave function where 99\% of density lies within 5 $\AA$. (b) The $\delta E_b$ comparison of model prediction by using Eq.~\ref{deltaEb}, Eq.~\ref{Wmodel}, and Eq.~\ref{DeltaW} with results from full \textit{ab-initio} BSE. Good agreement between BSE results (blue dots) and model prediction (orange line) is obtained in the relatively low substrate dielectric constant range. In this range, $\delta W$ varies slower, better satisfying the approximation in Eq.~12. 
    }
    \label{Fig7}
\end{figure}

We plotted the change of screened Coulomb potential, $\delta W(\rho,\epsilon_{sub})$ in Fig.~\ref{Fig7} (a), where the defect layer thickness $d=6.66 \AA$ is extracted from inter-layers distance, and defect layer dielectric constant $\epsilon_{def}$ is obtained in SI section V. 
%has been converted into dielectric at corresponding thickness(See SI section V). 
The dashed line at 5 $\AA$ represents the maximum spanning range of the localized defect states $1b_1$ and $2b_1$.

Fig.~\ref{Fig7}(b) shows that the $\delta E_b$ calculated by Eq.~\ref{deltaEb}, Eq.~\ref{Wmodel}, and Eq.~\ref{DeltaW} compares reasonably well with \textit{ab-initio} BSE calculations (blue dots), especially in low substrate screening regime. The underestimation of $\delta E_b$ at the high dielectric constant of substrates $\varepsilon_{sub}$ range may be due to an over-simplification of model $W$ in Eq.~\ref{Wmodel}.
%It overestimates the energy as the substrate dielectric constant increases. This overestimation occurs because the screened interaction potential at higher substrate dielectric screening can no longer be considered constant within the spanning range of the defect orbital wavefunction, %\textcolor{red}{causing the high localization assumption break}, as shown in Fig.~\ref{Fig7}(a). 
Nevertheless, the full \textit{ab-initio} GW/BSE calculation suggests that the cancellation effect for defect state transitions between exciton binding energy ($\delta E_b$) and GW quasiparticle energy renormalization ($\delta E_{GW}$) still holds even at high substrate dielectric screening. 

%Our Frenkel exciton model at the high localization limit can provide a qualitative understanding of the underlying physics of this cancellation effect, but accurate estimation of exciton binding energy renormalization still requires $ab-initio$ simulation or an analytic model with assumption beyond the high localization limit.

\section{Conclusions}
In this work, we have explored the impact of strain, layer thickness, and substrate effects on the electronic and optical properties of point defects in 2D insulator - hexagonal boron nitride (hBN). Our investigation takes into account the effects of many-body interaction and excited-state relaxation. We first analyzed various contributions to strain susceptibility of ZPL, 
and found the dominant contributions often stem from the changes at the single-particle level. We explained the ZPL shift direction under strain through molecular orbital theory, which is defect dependent, relying on the chemical bonding nature at the defect center.
 
%Furthermore, we provide an estimation of the zero-phonon line (ZPL) strain susceptibility and emphasize the importance of considering these effects in strain engineering, particularly with respect to the angular dependence. We have analyzed the local strain and wave function and identified the primary molecular orbitals that contribute to the variety of strain response behaviors in different defects and along different directions. 

Next our \textit{ab-initio} calculations demonstrate the robustness of optical peak position of defect single photon emitters (SPEs) when varying number of layers and substrates. We reveal the perfect cancellation between the renormalization of quasiparticle energy gap and exciton binding energy due to environmental screening.  To further understand this result, we derived analytical models based on solving BSE with Wannier basis for Frenkel excitons, to reveal the underlying mechanism and the required condition, i.e. the localization nature of defect bound exciton. We then use simple image-charge model for screened Coulomb potential to further validate such cancellation effect. 

%developed a Wannier basis model to unveil the underlying mechanism and condition for such optical peak robustness.

Our findings provide in-depth insights of mechanism and conditions that control the environmental impact on the properties of quantum defects, and shine light on the emerging research on  strain and substrate engineering of quantum defects in 2D materials.

\begin{acknowledgments}
We acknowledge the support by the National Science Foundation under grant no. DMR-2143233.
This research used resources of the Scientific Data and Computing center, a component of the 
Computational Science Initiative, at Brookhaven National Laboratory under Contract No. DE-SC0012704,
the lux supercomputer at UC Santa Cruz, funded by NSF MRI grant AST 1828315,
the National Energy Research Scientific Computing Center (NERSC) a U.S. Department of Energy Office 
of Science User Facility operated under Contract No. DE-AC02-05CH11231,
the Extreme Science and Engineering Discovery Environment (XSEDE) which is supported by National 
Science Foundation Grant No. ACI-1548562 \cite{townsXSEDEAcceleratingScientific2014}.
\end{acknowledgments}

%\nocite{*} %show all the bib, include those not cited 
\bibliographystyle{apsrev4-1}
\bibliography{Ref_all}

\end{document}

%\preprint{AIP/123-QED}
\beginsupplement

\title{Supplementary Materials: Effect of Environmental Screening and Strain on Optoelectronic Properties of Two-Dimensional Quantum Defects}% Force line breaks with \\

\author{Shimin Zhang}
\affiliation{Department of Physics, University of California, Santa Cruz, CA, 95064, USA}
\author{Kejun Li}
\affiliation{Department of Physics, University of California, Santa Cruz, CA, 95064, USA}
\author{Chunhao Guo}
\affiliation{Department of Chemistry and Biochemistry, University of California, Santa Cruz, CA, 95064, USA}
\author{Yuan Ping}
\email{yuanping@ucsc.edu}
\affiliation{Department of Chemistry and Biochemistry, University of California, Santa Cruz, CA, 95064, USA}

\maketitle

% define the color for Kejun's modification
\definecolor{kc}{rgb}{0.6,0,0.6}
\definecolor{del}{rgb}{0.66,0.66,0.66} %deleted part
\definecolor{sz}{rgb}{0.03,0.27,0.49} %Shimin's correction

%Convergence test of cbcn
\section{NUMERICAL CONVERGENCE OF GW/BSE CALCULATIONS}

\begin{figure}[h]
\centering
\includegraphics[width=\linewidth]{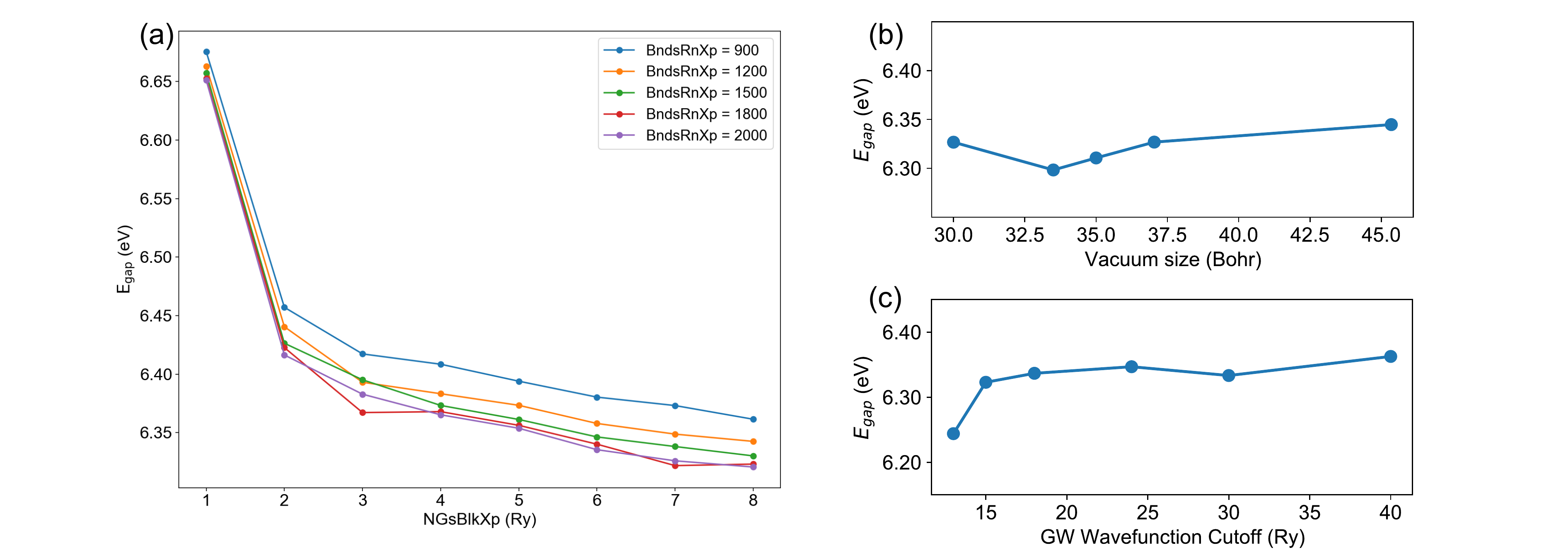}
\caption{GW convergence tests of HOMO-LUMO gap $E_{gap}$ of $\cbcn$ defect in hBN, with respect to (a) the number of bands (BndsRnXp) and response block size (NGsBlkXp) for the dielectric matrix, (b) vacuum size, and (c) energy cutoff of dielectric matrix and GW self-energy. 
}
\label{SI1}
\end{figure}

Fig.\ref{SI1} summarizes GW convergence tests using $\cbcn$ in hBN as an example. In Fig.\ref{SI1}(a) we show the band number BndsRnXp = 1800 and dielectric function block size of 8 Ry converge quasiparticle energies up to 50 meV. Fig. \ref{SI1}(b) shows that a vacuum size of 33.5 Bohr is sufficient, where the Coulomb truncation is utilized to accelerate the convergence \cite{rozziExactCoulombCutoff2006}. Lastly, the wave function cutoff energy of 30 Ry is chosen for GW self-energy as shown in Fig. \ref{SI1}(c).

\begin{figure}[h]
\centering
\includegraphics[width=\linewidth]{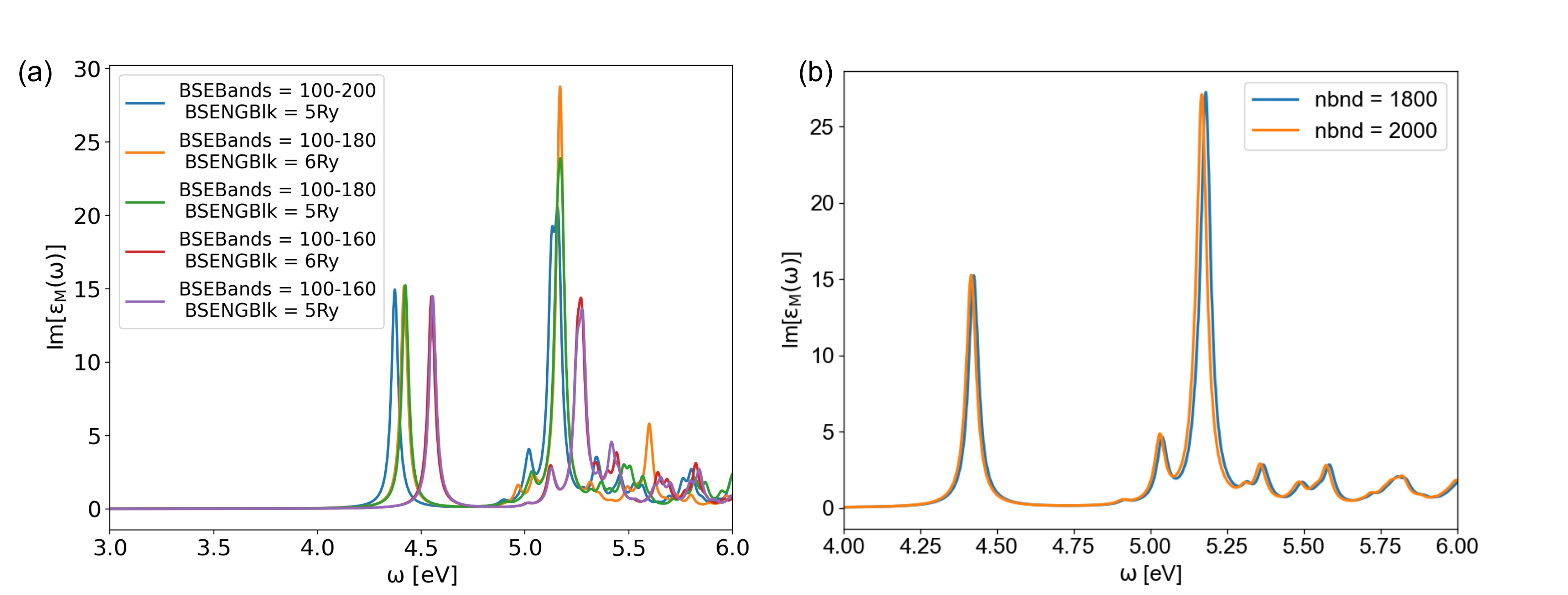}
\caption{The $\cbcn$ BSE spectrum convergence test of $\cbcn$, with respect to (a) bands used to construct the electron-hole pair wave function in BSE (BSEBands) and screened interaction block size (BSENGBlk), and (b) number of bands in GW level calculation for dielectric function (nbnd).}
\label{SI2}
\end{figure}

%\pagebreak
Fig.~\ref{SI2} (a) shows the BSE spectrum is converged at bands chosen for the BSE kernel from 100 to 180 (43 unoccupied and 37 occupied bands) and the BSE Kernel size of 5 Ry. Fig.\ref{SI2} (b) shows the BSE spectrum with respect to the total number of bands chosen for dielectric matrices in the screened Coulomb potential. 
%and confirmed 1800 bands is sufficient. 
%(here 144 is the LUMO band number) 

%BSE spectrum 
\section{BSE spectrum of $\cbcn$ and $\nbvn$ Defects in hBN}
\begin{figure}[h]
\centering
\includegraphics[width=\linewidth]{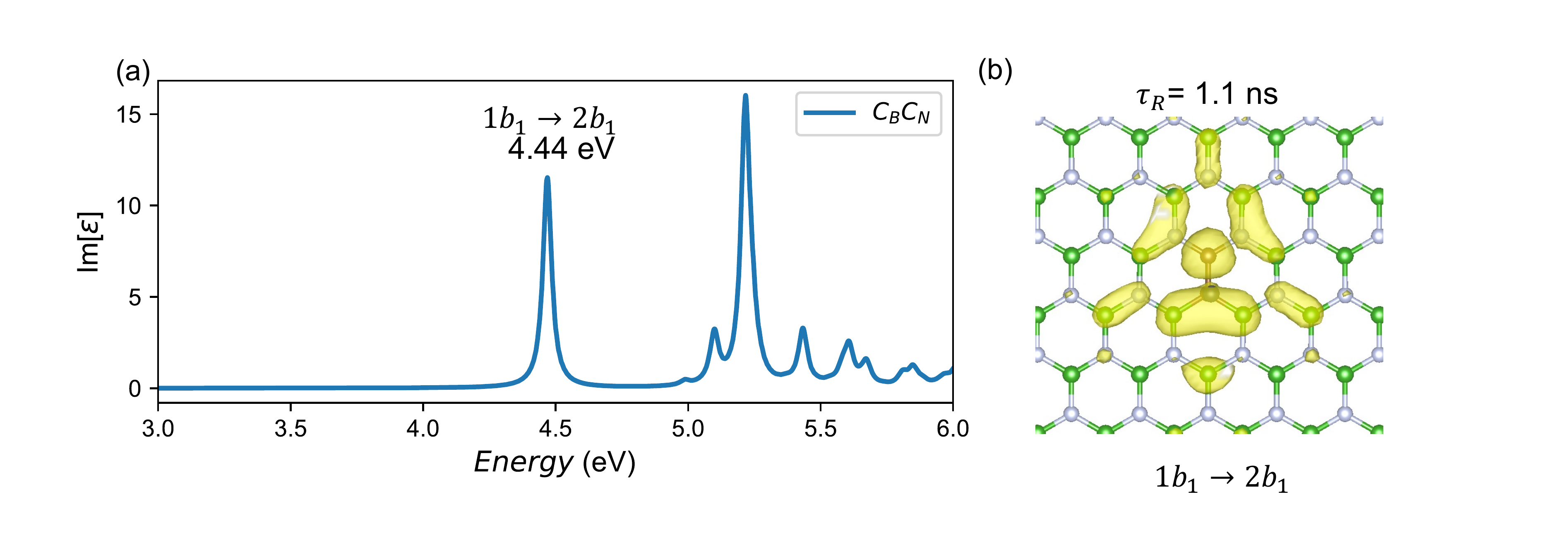}
\caption{(a) The BSE spectrum of $\cbcn$ in hBN. (b) The exciton wave function related to the defect transition $1b_1\rightarrow 2b_1$ in $\cbcn$, with an isosurface level of 0.005 $\AA^{-3}$.  }
\label{SI3}
\end{figure}

\textbf{$\cbcn$ defect in hBN} -  
Fig. \ref{SI3}(a) shows the BSE spectrum of $\cbcn$ in monolayer hBN, which has a sharp peak at 4.44 eV for defect state transition. The related exciton wave function is highly localized around the carbon defect center, as shown in Fig. \ref{SI3}(b).

\begin{figure}[h]
\centering
\includegraphics[width=\linewidth]{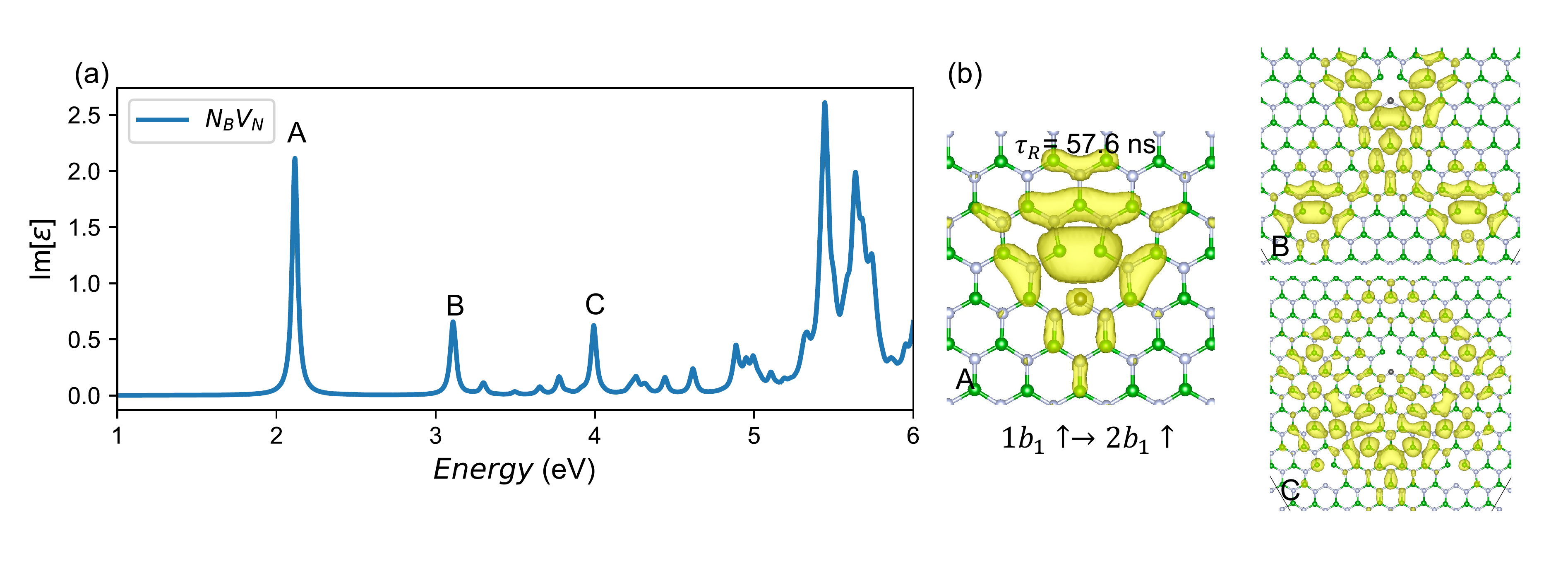}
\caption{(a) The BSE spectrum of $\nbvn$ with three defect-related transitions A, B and C. (b) The three exciton wave functions related to peaks A, B, C, with the hole fixed at the substitution Boron atom and an isosurface level of $0.015 \AA^{-3}$. }
\label{SI4}
\end{figure}

\textbf{$\nbvn$ defect in hBN} - 
Fig.\ref{SI4} shows the BSE spectrum and exciton wave functions of three defect-transition related peaks for $\nbvn$. The BSE spectrum in Fig. \ref{SI4}(a) labeled the three peaks as A,B,C, which are related to the $1b_{1\uparrow}\rightarrow 2b_{1\uparrow}$ transition. The main defect peak A, at 2.12 eV, is of primary consideration in the main text, and its exciton wave function solely includes $1b_{1\uparrow}\rightarrow 2b_{1\uparrow}$ transition. Peaks B and C are related to the mixing of defect-defect state transition and defect-bulk state transition. The exciton wave functions are plotted with the hole fixed at the substitutional Boron atom. Since the exciton wave function of peak B and C include delocalized bulk state component besides the defect states, they are less localized than peak A, and thus more sensitive to the change of environmental dielectric screening as we pointed out in the main text. \\

%Note that the intensity of the bulk peaks in the BSE spectrum shown here is underestimated because only a few tens of bulk bands are included in the BSE calculation. However, this is less relevant to the study of the defect peak position. \textcolor{red}{YP: plot smaller range, e.g. up to less than 6 eV, appropriate for band range }
%This underestimation is unrelated to the peak position as well as the emitter optical properties estimation.

\section{The low-symmetry structure and its optical transitions of $\nbvn$}

\begin{figure}[h]
\centering
\includegraphics[width=\linewidth]{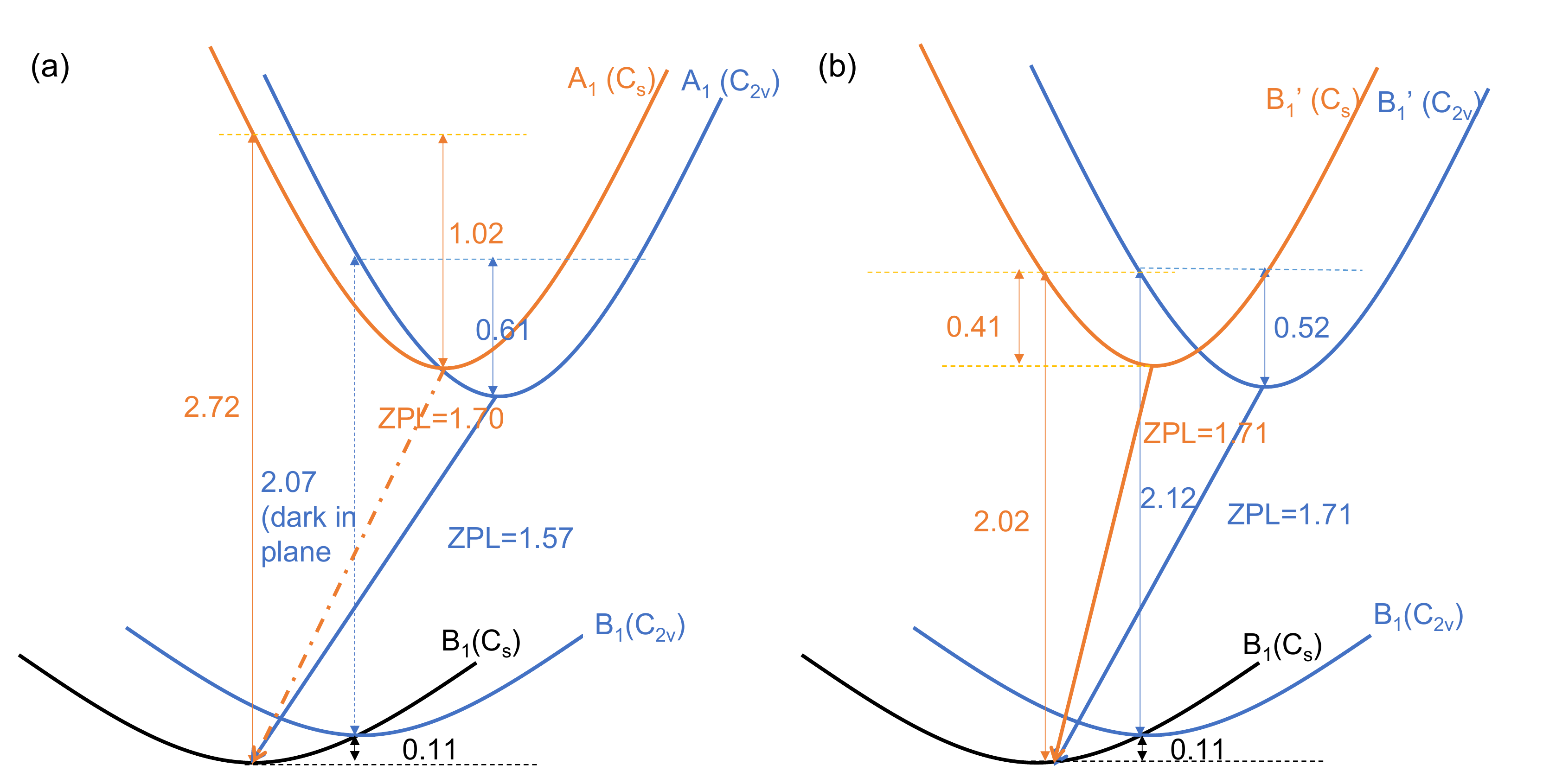}
\caption{The energy diagram of the transition from ground state $B_1$ to (a)$A_1$ and (b)$B_1^\prime$ excited state, where $C_s$ and $C_{2v}$ in the parentheses denote the symmetry of the state.}
\label{SI5}
\end{figure}

%We indeed realized the complexity of the geometry and transition for $\nbvn$ system; 
We investigated both in-plane ($C_{2v}$ symmetry) and out-of-plane distorted geometry ($C_s$ symmetry) while considering both possible excited states $A_1$ and $B_1^\prime$, which could account for the observed SPE. We took into account both the Frank-Condon shift and many-body effect, and summarized the energy diagram of two excited states and ground state under different symmetries in Fig.\ref{SI5}. 

Here we use multi-particle wave function label to denote the ground state and excited states. The $B_1$ is the ground state with the electron configuration of $[1a_1]^2[1b_1]^1[2b_1]^0$ as shown in the main text Fig.2. The excited state $A_1$ corresponds to the electron configuration of $[1a_1]^1[1b_1]^2[2b_1]^0$, and $B_1^\prime$ corresponds to the $[1a_1]^2[1b_1]^0[2b_1]^1$. 

\begin{figure}[h]
\centering
\includegraphics[width=\linewidth]{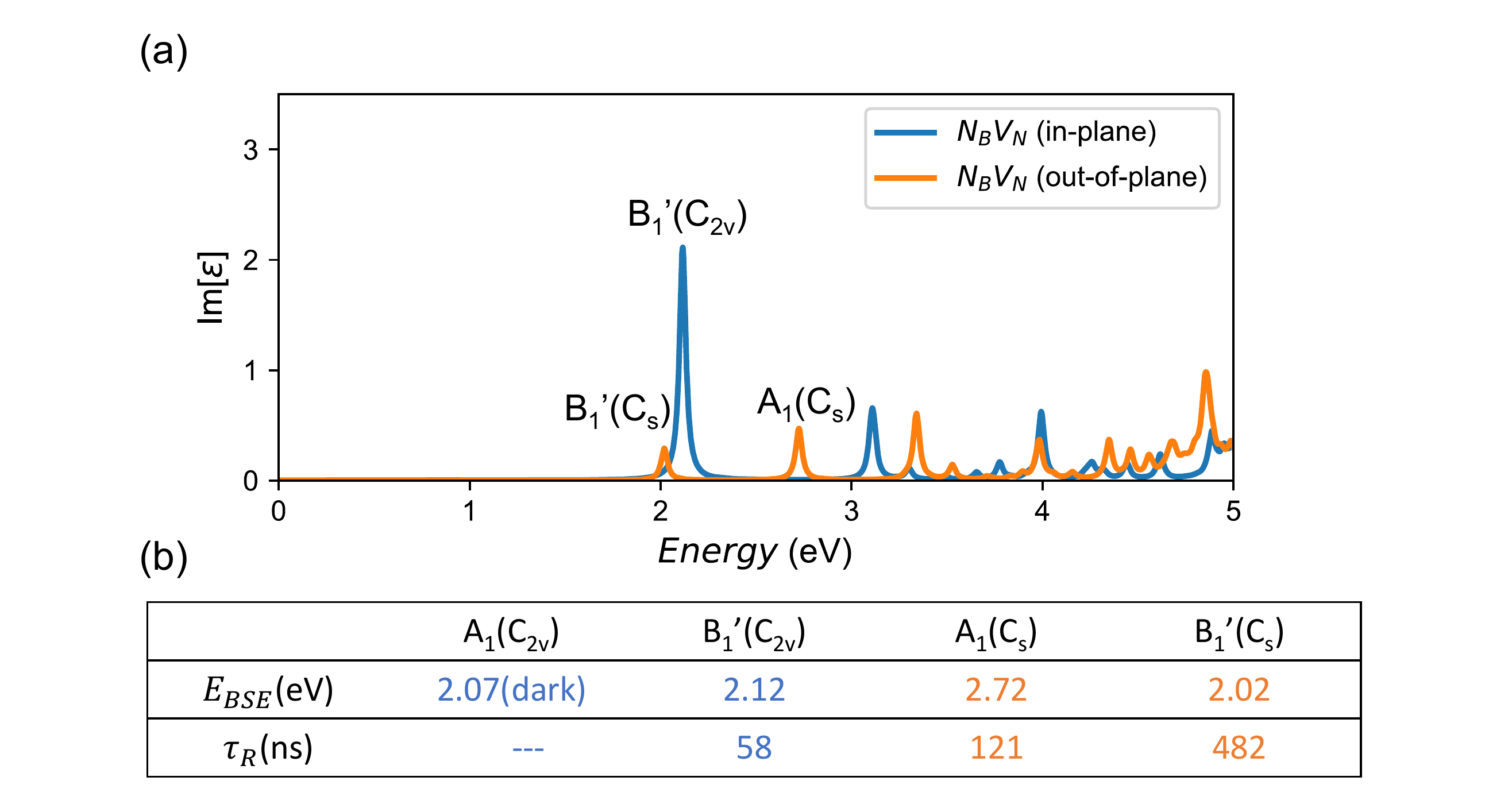}
\caption{(a) The BSE spectrum of $\nbvn$ with in-plane ($C_{2v}$ symmetry) and out-of-plane ($C_s$ symmetry) geometry. (b) The BSE excitation energy and corresponding radiative lifetime for both
structures.}
\label{SI6}
\end{figure}

Fig. \ref{SI5}(a) shows the energy diagram of the ground state $B_1$ and the excited state $A_1$, while Fig. \ref{SI5}(b) shows the ground state $B_1$ and the excited state $B_1^\prime$. The system is calculated under both $C_{2v}$ and $C_s$ symmetries, denoted in the parentheses, for all three states. The relaxation energy is calculated at the DFT level while the vertical excitation energy is calculated by GW/BSE, which is given in Fig. \ref{SI6}. The ground state energy is found to be 0.11 eV lower at $C_s$ symmetry than $C_{2v}$, with out-of-plane distortion of 0.66 $\AA$, consistent with the previous studies~\cite{liGiantShiftStrain2020}\cite{gaoRadiativePropertiesQuantum2021}. On the other hand, both $A_1$ and $B_1^\prime$ excited states have lower energy at $C_{2v}$ symmetry. This indicates that both excited states will experience a local structural distortion from $C_s$ to $C_{2v}$ when excited from the $C_s$ symmetry ground state. 

 The $B_1$-to-$A_1$ transition has been the main source of SPE, since it is the lowest transition and has been estimated to have a 2 eV ZPL that lies within the experimentally observed SPE range~\cite{abdiColorCentersHexagonal2018,liGiantShiftStrain2020}. Fig.~\ref{SI5}(a) shows that at $C_{2v}$ symmetry, the $B_1$-to-$A_1$ transition is forbidden, but it becomes allowed as the symmetry breaks to $C_s$ with a vertical transition energy of 2.7 eV. The details of the absorption spectrum are presented in Fig.~\ref{SI6}, which are consistent with previous calculations at the same level of theory and the same structural symmetry~\cite{gaoRadiativePropertiesQuantum2021}. The most viable optical excitation-recombination path for ZPL between $B_1$ and $A_1$ is to first excite from the $C_s$ symmetry ground state to $C_s$ symmetry $A_1$, undergo an $\sim$1 eV reorganization within $A_1(C_s)$, followed by local structural distortion to  $A_1(C_{2v})$, and then optically recombine to $B_1(C_s)$. Such a path has a long reorganization path and small ZPL of 1.57 eV (listed in Fig. \ref{SI6}). We believe it is unlikely to provide a bright SPE peak that falls within the experimentally observed SPE range of 1.6 to 2.2 eV.
%posses an 1.5 ms long radiative lifetime and thus be considered as forbidden compared to $B_1^\prime$, 

  We have chosen the $B_1$-to-$B_1^\prime$ transition as our primary focus in our study. As depicted in the energy diagram in Fig. \ref{SI5}(b), this recombination process involves a shorter relaxation path for the excited state. The $B_1^\prime$ excited state in the $C_s$ and $C_{2v}$ symmetry have similar relaxed energy, and therefore, can both contribute to an estimated ZPL of 1.7 eV. 

%Experiment
\section{Experiment result conversion}
In the experiment work we cited in main text the origin measurement of strain susceptibility is under the experimentally defined strain, which usually denote as the strain that is applied macroscopically on substrate~\cite{grossoTunableHighpurityRoom2017a} or applied by homogeneous pressure~\cite{xueAnomalousPressureCharacteristics2018}. Therefore we converted the experimental data for proper comparison.

In Grosso et al's work~\cite{grossoTunableHighpurityRoom2017a} the ZPL strain susceptibility was measured as -3.1 meV, 3.3 meV, and 6 meV/\% for three different emitters. These results are not directly comparable to our calculated strain susceptibilities, because their strain is applied and defined by the bending of the substrate, while our strain is defined by the stretching rate of defect layer itself. They introduced the strain transfer rate among two consecutive layers as t $\sim$ 98\%. The effective strain at the emitter location is $S^{N}_{eff}=S^{N-1}_{eff}\cdot t=\dots = \epsilon \cdot t^{N} $. For number of layers $N \sim 150$ in this experiment, the local strain can be 20 times smaller than applied strain, i.e. $S_{eff} = \epsilon\cdot 0.05 $. We apply this conversion relation to their experiment result and obtained the strain susceptibilities of -61 meV, 66 meV, and 120 meV/\% for the three emitters, which is of the same scale as our result.

In Xue et al's work~\cite{xueAnomalousPressureCharacteristics2018}, 9 different SPE between 1.7-2.2 eV are investigated with external pressure. The pressure-strain conversion table provided in this paper indicates a conversion relation of -0.049\%\ strain per 1 GPa pressure for in-plane strain. The converted strain susceptibility in TABLE.\ref{tab_exp1} is ranges from -287 meV/\%\ to 214 meV/\%. Although these values are for the biaxial strain, our calculated results fall within the appropriate scale.
\begin{table*}[!ht]
\caption{\label{tab_exp1} Measured data from experiment \cite{xueAnomalousPressureCharacteristics2018} }
\begin{ruledtabular}
\begin{tabular}{cccccccccc}
        ZPL(eV) & 2.125 & 2.1 & 2.065 & 2.035 & 2.025 & 1.985 & 1.925 & 1.9 & 1.76 \\ \hline
        Pressure coefficient (mev/Gpa) & -6.1 & 14.1 & -4.5 & 2.7 & -2.5 & -10.3 & -5.9 & -3.9 & -10.5 \\ \hline
        Strain susceptibility \footnote{biaxial strain} (meV/strain)  & 124.49  & -287.76  & 91.84  & -55.1  & 51.0  & 210.2  & 120.41  & 79.59  & 214.29 \\ 
\end{tabular}
\end{ruledtabular}
\end{table*}\\

\section{System Dielectric constant and 2D effective polarizability}
The Table~\ref{tab2} gives the dielectric constant and 2D effective polarizability of mono-layer(ML), two layers(2L), three layers(3L) $\cbcn$, as well as $\cbcn$ on $\sns2$ substrate($\cbcn|\sns2$) and $\mos2$ substrate ($\cbcn|\mos2$). In this table the $\epsilon_{sub}$ represents the dielectric constant of substrate or other layers, $\alpha_{2D,sub}$ is the 2D effective polarizability of substrate or other layers, and $\alpha_{2\text{D},tot}$ is the system total 2D effective polarizability. 
\begin{table}[h]
	\begin{ruledtabular}
	\begin{tabular}{cccccc}
      			   &ML & 2L & 3L & $\cbcn|\sns2$ & $\cbcn|\mos2$ \\
	\hline
	$\alpha_{2\text{D},tot}$    &   1.946  &    3.814     & 5.682        &     8.421   &  15.009\\
	$\epsilon_{sub}$ 	    & 1	    & 1.589        &  2.020        &     3.047   &  5.125 \\
	$\alpha_{2\text{D},sub}$ & 0    & 1.868       & 3.735      &    6.475         &  13.063   \\
	\end{tabular}
	\end{ruledtabular}
\caption{\label{tab2} System total 2D effective polarizability $\alpha_{2\text{D},tot}$, substrate dielectric $\epsilon_{sub}$, and substrate 2D effective polarizability $\alpha_{2\text{D},sub}$ of each $\cbcn$ system.}
\end{table}

The dielectric constant of 2D systems depends on the lattice length along the out-of-plane direction, denoted as $L$. We calculated the dielectric constant of the monolayer defective hBN system at $L=30.00$ Bohr to be $\epsilon_{def}(L=30.00)=1.815$ at DFPT. In the image charge model for the screened potential change, we set the defective hBN monolayer thickness to $d=12.59$ Bohr, twice of the bulk hBN interlayer distance. To obtain the corresponding $\epsilon_{def}(L=12.59)$, we use the 2D effective polarizability as the bridging quantity, which is independent on the lattice length or vacuum sizes, i.e. $\alpha_{def}(L=30)=\alpha_{def}(L=12.59)$. Combining this with the relation between 2D effective polarizability and dielectric constant, $\alpha_{2\text{D},def}(L)=(\epsilon_{def}-1)\cdot L/(4\pi)$, we can determine the dielectric constant at $L=12.59$ to be $\epsilon_{def}(L=12.59)=2.940$, which is used in our model calculations in the main text.

\bibliography{Ref_all}